\documentstyle[11pt,psfig,aaspp4]{article}  
\lefthead{H.W. Duerbeck et al.}
\righthead{The rise and fall of V4334 Sgr}

\begin{document}

\title{The rise and fall of V4334 Sgr (Sakurai's Object)}

\author{\sc H.W. Duerbeck\altaffilmark{1}, 
W. Liller\altaffilmark{2}, 
C. Sterken\altaffilmark{1}, 
S. Benetti\altaffilmark{3}, 
A.M. van Genderen\altaffilmark{4}, 
J. Arts\altaffilmark{4}, 
J. Kurk\altaffilmark{4}, 
M. Janson\altaffilmark{5}, 
T. Voskes\altaffilmark{4},
E. Brogt\altaffilmark{5}, 
T. Arentoft\altaffilmark{1}, 
A. van der Meer\altaffilmark{4},
and R. Dijkstra\altaffilmark{5}}
\altaffiltext{1}{WE/OBSS, Free University Brussels (VUB), 
Pleinlaan 2, B-1050 Brussels, Belgium, e-mail: hduerbec, csterken, 
tarentof@vub.ac.be}
\altaffiltext{2}{Instituto Isaac Newton, Ministerio de Educacion de Chile, 
Casilla 8-9, Correo 9, Santiago, Chile, e-mail: wliller@compuserve.com}
\altaffiltext{3}{TGN, Centro Galileo Galilei, Calle Alvarez de Abreu, 70, 
E-38700 Santa Cruz de La Palma, Canary Islands, Spain, 
e-mail: benetti@tng.iac.es}
\altaffiltext{4}{Leiden Observatory, Postbus 9513, NL-2300 RA Leiden, 
The Netherlands, e-mail: genderen@strw.LeidenUniv.nl}
\altaffiltext{5}{Kapteyn Institute, University of Groningen, P.O. Box 800, 
NL-9700 AV Groningen, The Netherlands}

\small

\begin{abstract}

CCD {\it UBVRi\/} photometry of the final helium flash object V4334 Sgr 
(Sakurai's Object), carried out during 1997 -- 1999, is presented, 
and the light curve from its pre-discovery rise to the dust obscuration 
phase is constructed. The optical light curve can be divided into 
four sections, the rise to maximum, the maximum, the dust onset, 
and the massive dust shell phase. The color indices show a general 
increase with time, first because of the photospheric expansion and 
cooling, and later because of the dust forming events. The energy 
distributions for the years 1996 -- 1999 show that an increasing 
part of the energy is radiated at infrared wavelengths. In 1996, 
the infrared excess is likely caused by free-free radiation in the 
stellar wind. Starting from 1997 or 1998 at the latest, carbon 
dust grains are responsible for the more and more dramatic decrease 
of optical radiation and the growing infrared excess. Its photometric 
behavior in 1998 -- 1999 mimics the ``red declines'' of R CrB 
variables, the amplitude, however, is more extreme than any fading 
ever observed in an R CrB star. Evidence is given that a complete 
dust shell has formed around V4334 Sgr. It therefore shows similarities 
with dust-forming classical novae, although 
evolving $\sim 20$ times more slowly. Its luminosity increased by a 
factor 4 between 1996 and 1998. A comparison of time scales of the 
final helium flash objects FG Sge, V605 Aql and V4334 Sgr shows 
that the observed photometric and spectroscopic features are similar, 
while V4334 Sgr is the most rapidly evolving object to date.

\end{abstract}


\section{Introduction}
\label{intro}

After completing their core helium-burning phase, stars less massive than 
$\sim 10.5~M_\odot$ develop in their AGB phase electron-degenerate cores 
of carbon and oxygen, or oxygen and neon at the massive end, and alternately 
burn helium or hydrogen in shells. Each quiescent helium-burning phase is 
preceded by a thermonuclear runaway in the degenerate helium layer. In the 
aftermath of each of these ``thermal pulses'',  carbon and s-process 
elements are transported to the stellar surface (Iben \& MacDonald 
1995)\markcite{ibe95}. Finally, these stars undergo extensive mass loss 
(``superwind phase''), and move in the Hertzsprung-Russell diagram to the 
region of central stars of planetary nebulae (PNN). 

The peculiar variable star FG Sge, which is situated in the center of a 
planetary nebula, has inspired theoreticians to study the post-AGB 
evolution in more detail. It is now believed that in about 10\% of all 
thermally pulsing stars, the last pulse can occur in a very late stage, 
when the star has already settled down as a PNN. In such a case, the last 
pulse is directly observable as a ``final He flash'', which drives the 
star from the region of the PNN back to the top of the AGB. This  
happened to FG Sge in the course of the 20th century. Such a ``born-again 
giant'' phase can last decades, centuries or millennia, but the object will 
return finally to the PNN region. During the final He flash phase, 
the outer layers of the star undergo extensive nucleosynthesis, including 
more or less complete processing of the surficial hydrogen. A large 
fraction of the outer layers is ejected, leading to the formation of a 
hydrogen-poor, carbon-rich nebulosity in the center of the planetary 
nebula. 

Planetary nebulae with central hydrogen-poor condensations like Abell 30 or 
Abell 80 (Jacoby 1979\markcite{jac79}, Jacoby \& Ford 1983\markcite{jac83}), 
H-deficient post-AGB stars like Wolf-Rayet type central stars of planetary 
nebulae, or white dwarfs of the PG1159 type (e.g. Werner et al. 
1999\markcite{wer99}) are possible end-products of final He flash objects.

The evolution of a final He flash from the PNN to the giant stage may 
take decades (as in FG Sge) or only a few years, as concluded 
in recent years from the present state and a few historical observations 
of ``Nova Aquilae No. 4'' of 1919, also known as V605 Aql 
(Seitter 1985\markcite{sei85}, Clayton \& De Marco 1997\markcite{cla97}). 
77 years after the flareup of V605 Aql, a ``novalike object in Sagittarius'' 
was soon recognized as another one of these rare events. It offers 
the first opportunity to study in detail the evolution of a fast final 
He flash.

Sakurai's object, later named V4334 Sgr, was discovered on 1996 February 
20 as a star of $11^{\rm th}$ magnitude by Yukio Sakurai (Nakano, 
Benetti \& Duerbeck 1996\markcite{nak96}). Prediscovery observations by
Y. Sakurai and K. Takamizawa showed that it had been at magnitude 12.5  
in early 1995, and possibly at magnitude 15.5 in late 1994. Several 
groups have monitored the optical brightness evolution: a Russian group 
(Arkhipova \& Noskova 1997\markcite{ark97}, Arkhipova et al. 
1998\markcite{ark98}, 1999\markcite{ark99}), a US group using automatic 
photometric telescopes (Margheim, Guinan \& McCook 1997\markcite{mar97},  
Guinan et al. 1998\markcite{gui98}), and a Chilean-European group,
whose results are presented here. Furthermore, scattered observations 
made with larger telescopes during interesting phases of its evolution 
will also be discussed.
 
The {\it UBVRiz\/} observations made in 1996 by the Chilean-European group 
were published (Duerbeck et al. 1997\markcite{due97}, hereafter quoted as 
D97). In the present paper, photometric observations for the years 1997, 
1998, and 1999 are reported, covering the complete rise and decline of 
the brightness of V4334 Sgr in the optical region (Sect.~\ref{obs}). 
Its properties before the outburst are investigated and constraints are 
given on its distance (Sect.~\ref{prehistory}). The light and color curves 
are constructed and interpreted (Sect.~\ref{optical}). The observed fadings 
are explained by dust formation, and similarities and differences of these 
events with fadings of R CrB stars are outlined (Sect.~\ref{dust}). The 
development of the energy distribution between 0.36 and 15 $\mu$m is 
investigated, and parallels to dust-forming classical novae are shown 
(Sect.~\ref{energy}). Finally, the time scales of the final He flashes 
in V4334 Sgr, V605 Aql and FG Sge are compared, and predictions for their 
future evolution are given (Sect.~\ref{time}).

\section{Observations}
\label{obs}

{\it UBVRi\/} observations of V4334 Sgr were carried out with the 0.91 m 
Dutch light collector at ESO La Silla, until its shutdown on April 1, 1999. 
Filters and comparison stars were the same as those used in D97 (the 
filters $R$ and $i$ refer to Cousins $R_C$ and Gunn $i_G$, respectively). 
The object was also observed until 1998 October with the 0.2 m 
$f/1.5$ Schmidt telescope of W.~Liller in Re$\rm \tilde{n}$aca, 
Vi$\rm \tilde{n}$a del Mar, Chile. Additional observations were 
obtained with the 3.5 m Telescopio Nazionale Galileo (TNG)\footnote
{The Italian Telescopio Nazionale Galileo (TNG) is operated on the
island of La Palma by the Centro Galileo Galilei of the CNAA (Consorzio
Nazionale per l'Astronomia e l'Astrofisica) at the Spanish Observatorio
del Roque de los Muchachos of the Instituto de Astrofisica de Canarias}
in July, 1999.

The early observations with the Dutch telescope were analyzed using 
ROMAFOT aperture photometry. The continuing decline in brightness, 
especially at short wavelengths, made it necessary to carry out DAOPHOT 
profile fitting photometry of V4334 Sgr from 1997 September 15 onward.
Because of the increase of exposure times, comparison star (1) of D97 
became too bright to be used as a local standard, and the average magnitude 
of stars $(2)-(6)$ (taken from D97) was used as the reference magnitude 
in all filters. The Re$\rm \tilde{n}$aca observations are based on 
CCD aperture photometry. Since they were always taken relative to 
star (1) through a non-standard $V$ filter which extends towards the 
red, a transformation was established from simultaneous Dutch ($V,i$) 
and Re$\rm \tilde{n}$aca ($\Delta V^\star$) observations, which permits 
to convert the Re$\rm \tilde{n}$aca observations into the 
standard $V$ system:
$$
V = 10.65 +0.93\Delta V^\star +0.047(V-i),
$$
where $\Delta V^\star$ is the magnitude difference relative to the 
bright comparison star; the color $V-i$ is taken from the ``Dutch'' 
observations, taken near the time of the Re$\rm \tilde{n}$aca 
observations. The {\it UBVRi\/} data of 1997 -- 1999 are listed in 
Table~\ref{tab1}. 

\placetable{tab1}

\section{V4334 Sgr before its final He flash}
\label{prehistory}

V4334 Sgr before its final He flash was a faint blue star on the ESO/SERC 
Sky Atlas (Duerbeck \& Benetti 1996\markcite{due96a}, hereafter quoted as 
D96). In this section, the apparent magnitude is determined, the interstellar 
reddening is estimated, and upper and lower limits of its distance are given. 
This information is important for the derivation of the luminosity during 
the final He flash phase.

\begin{figure}
\psfig{figure=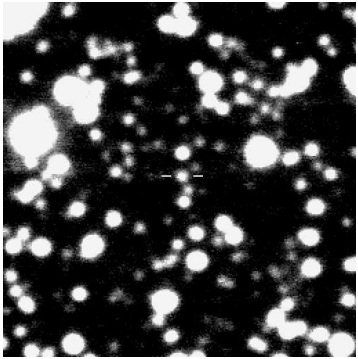,height=7cm}
\figcaption[duerbeck.fig1.ps]{The field of V4334 Sgr, taken with the 3.5 m
TNG at La Palma through an $R$ filter in July, 1999. The object is marked.
Its brightness is $R=20\fm 3$. The size of the field is $40'' \times 40''$.
North is top, east to the left.
\label{fig1}}
\end{figure}

Deep images with the TNG telescope in 1999 July allow to establish a 
preliminary faint magnitude scale in the vicinity of V4334 Sgr 
(Fig.~\ref{fig1}). A northern and a southern visual companion, each 
$2\farcs 5$ from V4334 Sgr, are seen on the Sky Atlas images (see plate 1 
of D96). The northern companion has $V=20.85, R=19.93$, the southern one 
$V=21.12, R=20.30$. Due to the lack of $B$-magnitudes in the TNG observations, 
pseudo-$B$-magnitudes were assigned to some stars with the help of the 
available $U$ and $V$ magnitudes. Using this preliminary scale, 
the visibility of some fainter field stars was checked on the Sky 
Atlas plates, and the pre-outburst photographic magnitudes of V4334 
Sgr were estimated to be $m_B\approx 21^{\rm m}$, $m_R > 21\fm 5$. 

\placefigure{fig1}

The interstellar reddening of V4334 Sgr is still poorly known. While 
$E_{B-V} = 0.54$ was estimated by D96, the value $0.71\pm 0.09$ was derived 
by Pollacco (1999)\markcite{pol99} from the observed H$\alpha$/H$\beta$ 
line ratio of the surrounding planetary nebula for case B. Another 
value, 1.15, was suggested by Eyres et al. (1998b)\markcite{eyr98b}.
Kimeswenger \& Kerber (1998)\markcite{kim98} made a detailed study of 
interstellar reddening in the field around V4334 Sgr and found $E_{B-V} 
= 0.90\pm 0.09$ for 18 stars with distances $d \ge 2.0$ kpc. Since we 
have reasons to believe that V4334 Sgr has a distance $d \ge 2$ kpc (see 
Sect.~\ref{energy}), we adopted $E_{B-V} = 0.8$, which is compatible with 
both Pollacco's and Kimeswenger \& Kerber's results. 

The brightness of the pre-outburst magnitude of V4334 Sgr was compared
with other PNNs. Absolute $B$ magnitudes of central stars of planetary 
nebulae were derived using information on trigonometric parallaxes 
given in Jacoby, De Marco \& Sawyer (1998)\markcite{jac98} and Acker 
et al. (1998)\markcite{ack98}. Apparent magnitudes were taken from the 
catalogue of Acker et al. (1992)\markcite{ack92}, the interstellar 
extinction was calculated with the model of Hakkila et al. 
(1997)\markcite{hak97}. Our sample is restricted to central stars 
in roundish bright and faint nebulae, while central stars inside 
irregular nebulae and stars which are obviously blended with field stars
were not considered. The derived absolute magnitudes are listed 
in Table~\ref{tab2}. Using the average value and its $1~\sigma$ 
deviation, $M_B = 6.4\pm 1.2$, and taking $m_B=21^{\rm m}$ for the 
pre-outburst magnitude of V4334 Sgr, its distance is derived to 
$1800^{+1400}_{-800}$ pc, assuming a reddening $E_{B-V}=0.8$. 
The value $E_{B-V} = 0.7$ would increase the distance estimates 
by 20\%. Jacoby et al. (1998)\markcite{jac98} concluded that the 
distance remains poorly determined, with possible values lying 
in the range 1 to 4 kpc. This is in agreement with the present 
result. This issue will be taken up again in Sect.~\ref{energy}.

\placetable{tab2}

\section{The optical outburst behavior of V4334 Sgr}
\label{optical}

\subsection{The light curve}
\label{vcurve}

\begin{figure}
\psfig{figure=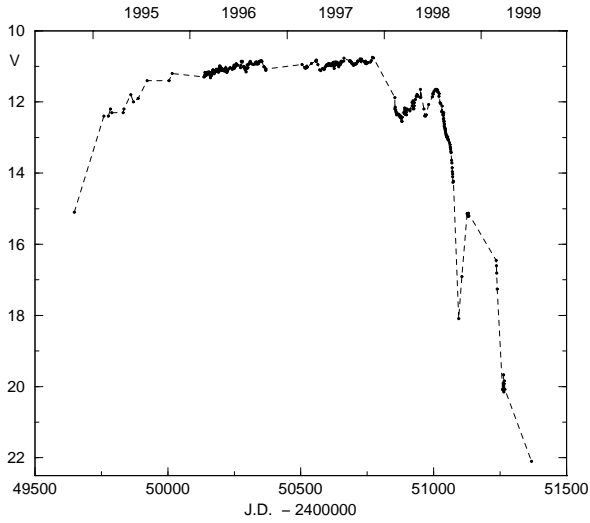,height=7cm,angle=270}
\figcaption[duerbeck.fig2.ps]{The $V$ light curve of V4334 Sgr, 1995 --
1999. The data of 1994-5 are the photographic pre-discovery observations 
of Takamizawa, converted to visual magnitudes, assuming that the
object is expanding at constant luminosity. Some observations by
Jacoby \& De Marco (1998) and Jacoby (1999) are also included.
\label{fig2}}
\end{figure}

Figure~\ref{fig2} shows the complete $V$ light curve of V4334 Sgr. 
It can be divided into four characteristic stages, conveniently 
separated by the seasonal gaps when the star was too close to the 
sun.

\placefigure{fig2}

The first stage is the ``rise to maximum'' of 1994 -- 1995, which is 
covered only by Takamizawa's 12 prediscovery observations (Takamizawa
1997\markcite{tak97}; see D97 for a discussion of the prediscovery light 
curve); note that the 1994 point is possibly only an upper limit.
The second one is the ``maximum stage'' of 1996 and 1997, when 
quasiperiodic brightness fluctuations were superimposed on an 
almost constant $V$ magnitude of $\approx 11^{\rm m}$.
The third one is the ``dust onset stage'' of early and mid-1998, 
when the object dropped in brightness by $1^{\rm m}$, and continued 
to show quasiperiodic fluctuations. Finally, the fourth one is 
the ``massive dust stage'' of late 1998 and 1999, when the object 
suffered dramatic declines of 3 to $11^{\rm m}$ in visible light, 
and when strong, erratic brightness fluctuations -- unfortunately 
poorly documented -- were present.

\begin{figure}
\psfig{figure=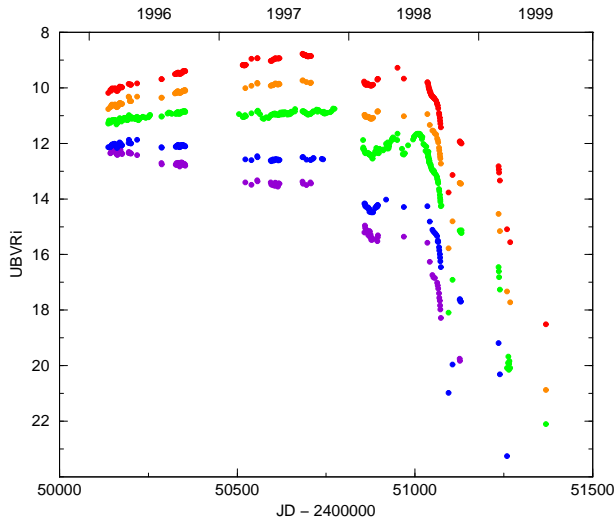,height=7cm,angle=270}
\figcaption[duerbeck.fig3.ps]{The {\it UBVRi\/} light curve of V4334 Sgr.
This curve includes only the observations made at the Dutch 0.91 m telescope
and the Re$\rm \tilde{n}$aca 0.2 m telescope, plus some observations at 
late stages made by Jacoby, Jacoby \& De Marco, and Benetti. From bottom 
to top: {\it U, B, V, R, i\/} light curves.
\label{fig3}}
\end{figure}

As already shown in D97, the first and second stages can be explained 
by an object with a slowly growing photosphere or pseudo-photosphere, 
radiating almost at constant luminosity. In contrast to a stable 
photosphere, a pseudo-photosphere is formed in a optically thick wind, 
which is driven by radiation pressure from an object radiating near 
Eddington luminosity (see, e.g. Bath \& Harkness 1989)\markcite{bat89}. 
Such a behavior is found in classical novae at early outburst stages. 
Dynamical instabilities causing mass loss and dust formation have also 
been suggested for R CrB stars which radiate close to the Eddington 
limit (Asplund 1998)\markcite{asp98}. The photosphere of V4334 Sgr may 
be a true photosphere or a pseudo-photosphere, an attempt to decide 
between both cases will be made later (Sect.~\ref{energy}). The 
expansion and cooling of the photosphere is documented in the 
{\it UBVRi\/} light curves (Fig.~\ref{fig3}), which show a shift of 
the radiation maximum towards longer wavelengths at later times. 

\begin{figure}
\psfig{figure=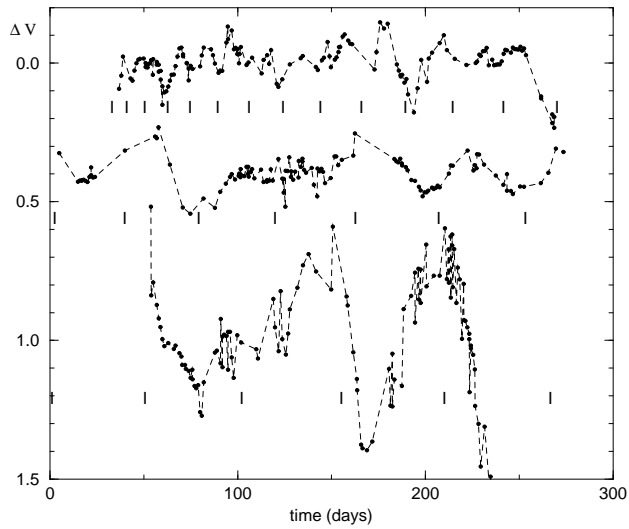,height=7cm,angle=270}
\figcaption[duerbeck.fig4.ps]{Brightness oscillations of V4334 Sgr, which are
superimposed on the {\it V\/} light curve. All available $V$ data have been
used for this diagram. The long-term brightness decline of the years 1996, 
1997 and 1998 (from top to bottom) has been removed to show the oscillations 
more clearly. The data of 1997 and 1998 were shifted by 0.4 and 0.9 mag 
to avoid overlaps. Times of brightness minimum as described by a formula 
of Arkhipova et al. (1999) are marked by vertical bars, placed below the 
corresponding light curve. 
\label{fig4}}
\end{figure}

\placefigure{fig3}
\placefigure{fig4}

Quasi-periodic or cyclic fluctuations are superimposed on this general 
photometric evolution. These variations can most easily be traced in 
the $V$ light curve, because the temporal coverage is highest, and because 
the effects of the temperature decline are least noticeable in {\it V}. 
Figure~\ref{fig4}, which also includes observations of the Russian 
group (see Sect.~\ref{intro}), shows the $V$ fluctuations in detail. 
For each of the years 1996, 1997, and 1998, the linear long-term trend 
in brightness was removed. In 1996, oscillations of short duration 
(26, 22 and 16 days and some shorter ones) are superimposed on a 70 
day oscillation. In 1997, a clear preference of a single oscillation 
with an amplitude of about $0.18\pm 0.03$ in $V$ and a period of 56 
days, prevailing over more than four cycles is seen. Secondary 
features are less important than in the year before (see also D97 
and Duerbeck et al. 1998)\markcite{due98}. 

The 1998 observations belong already to the third stage, which is 
influenced by dust. It is difficult to decide whether the fluctuations 
seen in early and mid-1998 are still caused by a pulsation or by dust 
obscuration events, especially since the multicolor coverage is poor. 
A characteristic time scale of $74\pm 8$ days is seen, but the 
periodicity is poorly defined, and masked by the strong brightness 
decline that occurred in the second half of 1998. The amplitude in 
$V$ has increased to $0\fm 75$.

In the fourth stage of late 1998 and of 1999, the object has faded 
dramatically. The poor temporal coverage, as well as the strong 
brightness fluctuations do not permit to study any underlying 
periodicities.

Summing up, the light curve between 1996 and mid-1998 can be described 
as being superimposed by quasiperiodic fluctuations of increasing cycle 
length and amplitude. Arkhipova et al. (1999)\markcite{ark99} claim 
that the variations can be described with a single period that increases 
linearly with time. Their ephemeris
$$
{\rm J.D. (min)}  = 2450133.1 + 6.91048\cdot E + 0.871788\cdot E^2
$$
was used to calculate the moments of minimum light, which are shown 
as vertical bars for the years 1996 -- 1998 in Fig.~\ref{fig4}. While a 
trend towards longer periods and larger amplitudes at later times is 
clearly present, the representation of minima by the above formula is 
not satisfactory, and the occurrence of both well-expressed and
marginal maxima and minima must be explained by the superposition of 
several pulsation modes, as was outlined in D87.

\subsection{V4334 Sgr in the two-color diagrams}
\label{twocol}

\begin{figure}
\psfig{figure=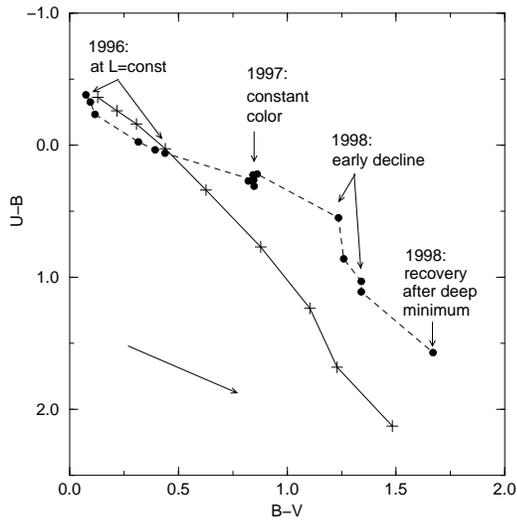,height=7cm,angle=270}
\figcaption[duerbeck.fig5.ps]{The $\ub$ vs. $\bv$ diagram of V4334 Sgr, 
1996 -- 1998, dereddened for the value $E_{B-V} = 0.8$ of interstellar 
reddening. The long arrow at the lower left shows the slope of the 
interstellar reddening line $E_{U-B}/E_{B-V}$. Important stages during 1996, 
1997, and 1998 are marked by arrows. In the maximum stage of 1996 -- 1997, 
the pseudo-photosphere cooled at constant luminosity (``at $ L=$ const''), 
but this cooling process seemed to have come to a halt in 1997 (``constant 
color''). The dust onset stage in early 1998 and the beginning of the 
massive dust stage in late 1998 are labelled here as ``early decline''. 
For the following brightness evolution, $U$ data are missing, except for a 
point taken during a temporary recovery of brightness in very late 1998
(``recovery after deep minimum''). The solid line marks the locus of 
hydrogen-deficient stellar atmospheres calculated by Asplund (1997),
where the effective temperatures of 9000, 
8500, 8000, 7500, 7000, 6500 and 6000 K are marked by plus-signs.
\label{fig5}}
\end{figure}

\begin{figure}
\psfig{figure=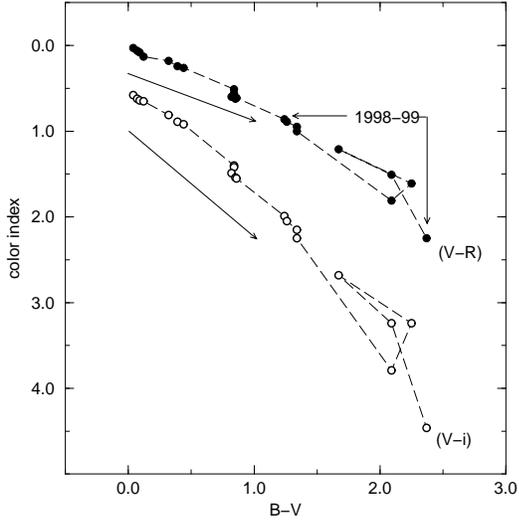,height=7cm,angle=270}
\figcaption[duerbeck.fig6.ps]{The $\vr$ vs. $\bv$ (filled circles) 
and the $V-i$ vs. $\bv$ (open circles) paths of V4334 Sgr, 1996 -- 1998, 
dereddened for $E_{B-V} = 0.8$. The two arrows in the upper left corner 
show the slopes of the interstellar reddening lines $E_{V-R}/E_{B-V}$ 
(upper arrow) and $E_{V-i}/E_{B-V}$ (lower arrow). 
The $V-i$ data points have been shifted by $+0\fm 5$ for clarity.
The Figure shows the gradual reddening of 1996, the stagnation of 1997, 
the subsequent dust formation of early 1998, as well as the loop of 
1998 -- 1999, marking the second fading, the subsequent recovery, 
and the third fading. 
\label{fig6}}
\end{figure}

The growth of the photosphere and the dust formation is 
studied in the multi-color light curve (Fig.~\ref{fig3}) and in 
two-color diagrams. The development of V4334 Sgr in the ($\ub$ 
vs.~$\bv$), ($\vr$ vs.~$\bv$) and ($V-i$ vs.~$\bv$) two-color diagrams 
is shown in Figs.~\ref{fig5} and \ref{fig6}. The colors are 
dereddened for $E_{B-V} = 0.8$. The observed averaged color indices 
for time intervals of days up to several weeks, and supplemented by 
observations of other authors at important phases, are given in 
Table~\ref{tab3}. 

\placetable{tab3}
\placefigure{fig5}
\placefigure{fig6}

The behavior of V4334 Sgr in the maximum stage is as follows. In 
1996, the star was continuously cooling. In Fig.~\ref{fig5}, it is 
moving along the two-color track of hydrogen-deficient carbon stars 
of hot to intermediate temperature, as calculated by Asplund 
(1997)\markcite{asp97}. This phase was interpreted by D97 as an 
object radiating at almost constant luminosity, while its photosphere 
is slowly moving outward with a velocity of $1~\rm km\,s^{-1}$. In 
1997, the star had moved away from the two-color track, and kept 
similar $\ub, \bv, \vr, V-i$ color indices for about 150 days, 
indicating that the photosphere had become stationary. In the
case of a pseudo-photosphere, whose location is determined by the 
actual mass-loss rate, this means that the mass-loss rate had 
stabilized.

\begin{figure}
\psfig{figure=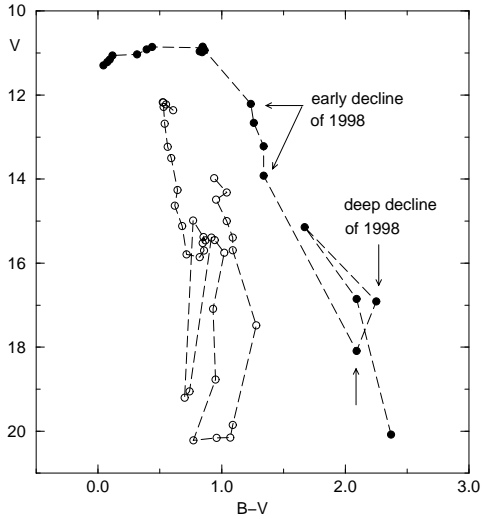,height=7cm,angle=270}
\figcaption[duerbeck.fig7.ps]{The $V$ vs. $\bv$ diagram of V4334 Sgr. The 
$\bv$ values are dereddened for $E_{B-V} = 0.8$. Two
points are available for the deep (second) decline of 1998, when the 
$V$-magnitude reached about $19^{\rm m}$, they are marked by arrows. 
After the brightness recovery in late 1998 and early 1999, V4334 Sgr 
dropped to fainter magnitudes and somewhat redder colors. 
The declines in brightness in 1998 and 1999 were similar to the  
``red'' declines of R CrB stars. For comparison, the deep decline 
of 1991 of the R CrB star V854 Cen, which is also composed of
several declines and recoveries, is shown (open circles; the $V$ magnitudes 
were shifted by $5^{\rm m}$ to fit into the diagram; data were taken 
from Lawson et al. 1992). 
\label{fig7}}
\end{figure}

During the dust onset and the massive dust stages, the colors show 
a reddening of increasing strength while the visible light of 
V4334 Sgr declined. The reddening increased noticeably in the deep decline 
of early October 1998, which is documented by observations of Jacoby 
\& De Marco (1998)\markcite{jac98}. Since no $U$ magnitudes were 
reported, this episode is missing from Fig.~\ref{fig5} and is only 
illustrated in the other two-color diagrams (Fig.~\ref{fig6}) and 
in the $(V, \bv)$ color-magnitude diagram (Fig.~\ref{fig7}). In 1999, 
the star had become so faint in $U$ that no observations were 
obtained. For 1999 July, an upper limit of $U = 23\fm 5$ is derived, but 
the previously recorded color indices make it likely that the star was 
several magnitudes fainter. Our $V$-observations show fluctuations 
around $19.5-20^{\rm m}$ in 1999 March, and a minimum brightness 
of $22\fm 1$ in 1999 July. Note that the very last $\vr$ and $V-i$ indices 
of 1999, which are not included in the Figures because of the lack
of $B$ magnitudes, indicate a decrease in reddening in spite of the 
faintness of the star (see also Sect.~\ref{dust}).

\placefigure{fig7}

\section{Dust forming events as seen in the optical region}
\label{dust}

Theoretical, spectroscopic and photometric results have pointed out a 
possible relation between R CrB stars and final He flash objects like FG Sge
(Gonzalez et al. 1998\markcite{gon98}, Jurcsik \& Montesinos 
1999\markcite{jur99}) and V4334 Sgr (Asplund et al. 1997\markcite{asp97},
Arkhipova et al. 1999\markcite{ark99}). 
Dust-forming events, similar to those seen in R CrB stars, have been 
suspected in the light curve of the rapidly evolving final He flash 
object V605 Aql (Harrison 1996)\markcite{har96}, and have been observed 
in recent years in the slowly evolving final He flash object FG Sge.  
Already in the first papers on V4334 Sgr, a dust-forming phase was 
predicted (D96, Duerbeck \& Pollacco 1996)\markcite{due96b}. The 
declines observed in V4334 Sgr (and first announced by Liller et al. 
1998a,b)\markcite{lil98a}\markcite{lil98b} can readily be compared 
with those observed in the other well-observed final He flash object 
FG Sge and in R CrB-type variables. Fading events of R CrB and V854 Cen, 
observed by Cottrell, Lawson \& Buchhorn (1990)\markcite{cot90}, 
Lawson \& Cottrell (1989)\markcite{law89} and Lawson et al. 
(1992)\markcite{law92} using multicolor photometry, were useful in 
comparing photometric and spectroscopic characteristics of He flash 
objects and R CrB variables.

\subsection{Pulsations and declines}
\label{puldec}

R CrB and related hydrogen-deficient stars show pulsations at maximum light. 
Periods are $40 - 100$ days, with amplitudes of a few $0\fm 1$. A tendency 
towards longer periods in cooler objects is seen (Lawson et al. 
1990)\markcite{law90}. In the R CrB star V854 Cen, 
the onset of a brightness decline usually occurs near maximum 
light of the pulsation cycle (Lawson et al. 1992\markcite{law92}), while in
RY Sgr, it occurs at minimum light (Menzies \& Feast 1997)\markcite{men97}.

The slowly evolving final He flash object FG Sge has shown pulsation periods 
ranging from 5 to 138 days, with a definitive trend to longer periods 
at later times, when the object had cooled (van Genderen \& Gautschy 
1995)\markcite{gen95}. After the first steep decline in 1992, FG Sge showed 
a nearly constant pulsation period of 115 days, and R CrB type fadings
often occurred near maximum light of the pulsation (Gonzalez et al. 
1998)\markcite{gon98}.

V4334 Sgr showed variations with cycles of $\approx 10 - 74$ days
with a tendency of better defined, longer periods at later (cooler)
stages, and amplitudes increasing from $0\fm 1$ to $0\fm 7$. 
Apart from the rapid change of behavior, the pulsations of V4334 Sgr are 
comparable to those of R CrB stars, as well as to the pulsations
of FG Sge. The three observed fading events of V4334 Sgr 
are separated by intervals of $\sim 200$ days, and do not seem to be 
related to the period of stellar pulsation. The well-documented second 
fading of 1998 September began around a minimum phase of pulsation.

\subsection{Color evolution during declines}
\label{colevo}

R CrB stars show both ``blue'' and ``red'' declines (Cottrell, Lawson 
\& Buchhorn 1990)\markcite{cot90}. The blue events presumably occur when 
the obscuring cloud is smaller than the photosphere, as seen from the 
observer. In a red decline of R CrB, the star moved along the line 
$(\ub)/(\bv) \approx 1$ in the two-color diagram. A well-observed red 
decline of V854 Cen had $(\ub)/(\bv) \approx 0.6$, $(\vr)/(\bv) \approx 
0.8$,  $(V-I)/(\bv) \gtrsim 1.6$.

The slowly evolving final He flash object FG Sge has only shown ``blue'' 
declines (Jurcsik \& Montesinos 1999)\markcite{jur99}. Thus, dust 
formation in FG Sge has been patchy until now.

The color behavior of V4334 Sgr is complex, because the growth of the 
photosphere, the dust formation, and even the interstellar reddening 
cause similar effects in two-color diagrams. While interstellar reddening 
just produces a constant shift in the diagram, the first two effects can 
most easily be disentangled with the aid of the $V$ vs.~$\bv$ diagram 
(Fig.~\ref{fig7}). It shows that the initial dust formation episode 
in the line of sight did not begin until early 1998.

V4334 Sgr shows a ``composite'' of several red declines, indicating that 
in all cases the whole visible photosphere is obscured (Fig.~\ref{fig2}). 
The first decline occurred at or before J.D. 2450853 (1998 February), 
and the star did not regain its former brightness; the second occurred 
around J.D. 2451045 (1998 September), and the star partly recovered; 
the third one occurred at or before J.D. 2451234 (1999 February), and 
after some fluctuations at $20^{\rm m}$, another decline followed, 
whose color characteristics are only poorly known. The value of 
$(\ub)/(\bv)$ appears to be always larger than in the case of R CrB 
stars. Following the onset of dust formation in 1998 February, a sequence of 
three normal points indicates the beginning of the fading of 1998 September
(this is marked ``early decline'' in Fig.~\ref{fig5}). After 
the deep decline in 1998 October, and the recovery in November, which 
was observed by Jacoby \& De Marco (1998)\markcite{jac98}, another 
$\ub$ color index is available (marked ``recovery after deep minimum''). 
The overall slope of the 1998 $(\ub)/(\bv)$ data is $\approx 2.4$.
This slope is substantially steeper than the slopes $\approx 1$ 
and $\approx 0.6$ observed in R CrB and V854 Cen. Whether this is
caused by different dust properties or a different amount of 
``chromospheric emission'' above the dust in these objects, cannot 
be decided on the basis of the available data. 

The other color index ratios, $(\vr)/(\bv) = 1.0$,  $(V-i)/(\bv) \approx
1.8$, derived for the interval 1998 -- 1999 when dust obscuration 
was obviously present, are surprisingly similar to those observed in 
V854 Cen (and, {\it cum grano salis}, to the interstellar reddening lines),
$(\vr)/(\bv) = 0.8$,  $(V-I)/(\bv) = 1.6$. Both V854 Cen and V4334 Sgr 
show a tendency to yield steeper slopes at very red colors and very 
faint magnitudes.

Even during a ``red'' decline, all color indices of V854 Cen turn to 
smaller values at very faint magnitudes. The fragmentary data of V4334 Sgr 
indicate that a similar effect occurred in 1999 July, when the $V$ 
magnitude reached its observed minimum near $22^{\rm m}$, and the
$\vr$ and $V-i$ indices were noticeably smaller than three months before.
Whether this is also the signature of an imminent brightness recovery, 
as it is found in R CrB stars, cannot be said because of lack of data.

\subsection{Depth and speed of declines}
\label{depspe}

The deepest decline in R CrB or related stars ever observed in the visible 
region was $8^{\rm m}$ (see the light curve of R CrB by Mattei, Waagen 
\& Foster 1991)\markcite{mat91}. Such a level may be reached during a 
single fading event or by a superposition of several fading events. 
In the latter case, the star does not become fainter, but may simply 
remain for a longer time at minimum level. Concerning the speed 
of declines, a ``red'' decline of R CrB, observed by Fernie, Percy 
\& Richer (1986)\markcite{fer86}, showed a maximum rate of decline of 
$0\fm 13~\rm day^{-1}$ in its later stages; a rate of $0\fm 27~\rm day^{-1}$ 
was observed by Cottrell et al. (1990)\markcite{cot90} during a ``blue'' 
decline. During the deep red decline of V854 Cen in 1991, with a 
superposition of three fading events, gradients up to $0\fm 7~\rm day^{-1}$ 
were observed (Lawson et al. 1992\markcite{law92}). 

During 1998 -- 1999, V4334 Sgr declined by $11^{\rm m}$ in $V$, i.e. 
the obscuration was at least an order of magnitude more efficient than
ever observed for an R CrB star. This decline consists of several 
superimposed fading events, as described in Sect.~\ref{colevo}. 
The object partially recovered from the first two events; the sparse 
data of 1999 indicate that V4334 Sgr is still obscured by the third
fading event (it is also possible that a fourth fading event took place). 
The first decline was not covered by observations, the decline rates of 
the second and third declines were $0\fm 05~\rm day^{-1}$ and 
$0\fm 14~\rm day^{-1}$. The speed and form of these declines resemble 
those of slow ``red'' declines of R CrB variables.

\subsection{Spectroscopic features of dust shells}
\label{spefea}

Dust that forms near an R CrB star experiences a strong radiation force,
moves outward and drags gas with it, which is collisionally excited.
R CrB, while recovering from a decline and at subsequent maximum light, 
showed a P Cyg line of \ion{He}{1} 10830 extending to $-240~\rm km\,s^{-1}$ 
(Querci \& Querci 1978)\markcite{que78}. 

In spectra of V4334 Sgr, a blueshifted absorption line of \ion{He}{1} 
10830 with an expansion velocity of $-550~\rm km\,s^{-1}$ was observed 
in 1998 March by Eyres et al. (1999)\markcite{eyr99}, i.e. shortly after 
the onset of dust formation the line of sight. No line had been present 
in spectra taken in 1997 July. From 1998 August onward, the line shape 
changed to a P Cygni profile (Eyres et al. 1999\markcite{eyr99}, 
Tyne et al. 1999)\markcite{tyn99}. During the deep minimum of 1999, 
the \ion{He}{1} line was observed in emission only, and extended less 
to the red than the P Cyg line of the previous year. It showed a 
blueshift of about $-500~\rm km\,s^{-1}$ relative to the radial 
velocity of V4334 Sgr (limits $-700$ and $+130~\rm km\,s^{-1}$). 
This indicates that (a) collisionally excited gas exists since early 1998, 
in agreement with the photometrically observed onset of dust formation; 
(b) while the stellar background faded, the line kept its strength 
and appeared as a P Cyg line in a semi-transparent shell; (c) after the 
massive dust formation, emission originating in the region moving away 
from the observer is almost completely obscured by dust, and only the 
emission originating in the hemisphere facing the observer is seen.

\subsection{The infrared behavior}
\label{infbeh}

The infrared behavior of R CrB stars correlates poorly with dust forming 
events in the line of sight, which mainly influence the flux in the optical
region. Since the dust cloud is small, it converts only a small fraction
of the total light of the star into infrared radiation. The infrared output
of an R CrB star is dominated by radiation from the overall circumstellar 
dust shell, which is heated by the star (and also shows the pulsational 
light variations observed in the star). On the other hand, the dust flux 
does not change significantly when the star goes into an obscuration
minimum. These findings are the best evidence for the 
patchiness of dust formation in the atmospheres of R CrB stars (Forrest,
Gillett \& Stein 1972\markcite{for72}, Feast et al. 1997\markcite{fea97}). 

The descent to minimum in V4334 Sgr was accompanied by a complete change
in its energy distribution, including a dramatic increase at infrared 
wavelengths (especially in the poorly observed range $\rm \lambda > 5~\mu m$).
A more detailed study of the steady growth of the infrared excess is 
given in Sect.~\ref{energy}.

\subsection{V4334 Sgr and R CrB stars: concluding remarks}
\label{concl}

Summing up, the fading events of V4334 Sgr in the years 1998 -- 1999
show striking similarities with the ``red'' declines of R CrB stars.
While there are differences between individual stars in the $\ub/\bv$ ratio,
all objects show a similar behavior at longer wavelengths.

In contrast to R CrB stars, the unusually deep, long-lasting minimum, 
the evolution of the \ion{He}{1} 10830 line structure, as well as the 
connection between optical fading and infrared brightening, indicate 
the formation of a {\it complete} dust shell around V4334 Sgr. The 
behavior of V4334 Sgr in 1998 and later should not necessarily be
described as the ``R CrB phase'', as Arkhipova et al. 
(1999)\markcite{ark99} have done. R CrB stars form no complete dust shells. 
Its behavior shows as well striking similarities with dust-forming 
classical novae, as will be shown in Sect.~\ref{evoclas}.

\section{Energy distribution and luminosity}
\label{energy}

The {\it UBVRi\/} photometry presented in this study can be combined with 
infrared photometry to a study of the overall energy distribution, the 
character of the infrared excess, and the time variation of the luminosity. 
Infrared photometry is available during each season:
Feast \& Whitelock (1999)\markcite{fea99} for 20 well-distributed {\it JHKL}
data sets, obtained between early 1996 to late 1999; Kamath \& Ashok 
(1999)\markcite{kam99} for 1996 and 1997 in {\it JHK};
Fouqu\'e (in D96) for 1996 April  in {\it IJK};
Arkhipova et al. (1998)\markcite{ark98} for 1996 and 1997 in {\it JHKLM};
Kimeswenger et al. (1997)\markcite{kim97} for 1997 March in {\it IJK};
Kerber et al. (1999)\markcite{ker99} for 1997 and 1998 at 7 wavelengths 
between 4.5 and 12.0 $\mu$m, observed with ISOCAM of ISO;
Lynch et al. (1998)\markcite{lyn98} for 1998 March and May in $L'$, $M'$, $N'$;
Kaeufl \& Stecklum (1998)\markcite{kae98} for 1998 June in $N$;
Jacoby (1999)\markcite{jac99} for 1999 April at 1.083 and 2.230 $\mu$m;
Tyne et al. (1999)\markcite{tyn99} for 1999 April and May in {\it JHKLM};
Hinkle \& Joyce (1999) for 1999 September in {\it JHKLM}.

\begin{figure}
\psfig{figure=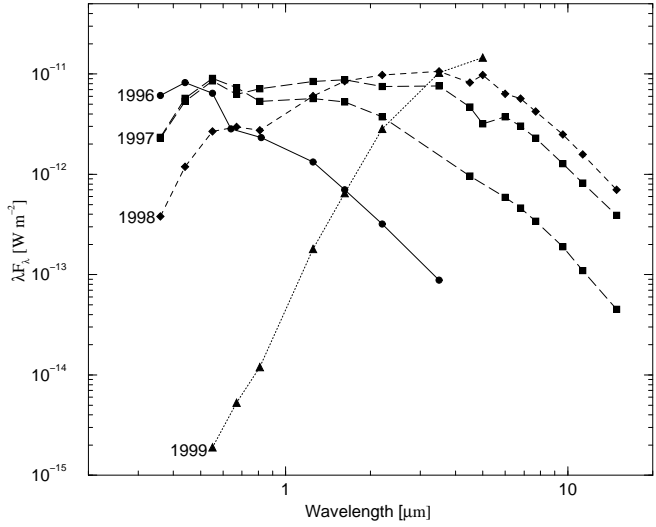,height=7cm,angle=270}
\figcaption[duerbeck.fig8.ps]{Selected energy distributions of V4334 Sgr 
between 1996 and 1999: 1996 March, circles, connected by lines; 1997 February
and September, squares, connected by long dashed lines; 1998 March: 
diamonds, connected by dashed lines; 1999 May: triangles, connected by
dotted lines. The stellar continuum curve is
easily recognized; its maximum shifts between 1996 and 1997 towards longer
wavelengths, between 1997 and 1998, such an effect is less noticeable.
Note that an infrared excess is always present, and that it increases 
in strength with time.
By 1998, it starts to influence the short-wavelength-region data, 
while in 1999, hardly an indication of the presence of the optical 
stellar continuum has remained.
\label{fig8}}
\end{figure}

A selection from these data was combined with quasi-simultaneous 
{\it UBVRi\/} data to construct twenty-one energy distributions of 
V4334 Sgr: seven for 1996, seven for 1997, three for 1998, and four 
for 1999. The magnitudes were dereddened for the value $E_{B-V} = 0.8$, 
and were converted into monochromatic irradiances $E_\lambda~(\rm in 
~W~m^{-2}~\mu m^{-1})$. They are listed in Table~\ref{tab4}. Integration 
over the irradiances $E_\lambda$ yielded total irradiances 
$(\rm in~W~m^{-2})$ which are also given. Selected results are shown 
in Fig.~\ref{fig8}.

\placetable{tab4}
\placefigure{fig8}

\subsection{The evolution of the infrared excess}
\label{evoinf}

Figure~\ref{fig8} shows that an excess of radiation at wavelengths 
$ > 1~\micron$ already exists during the earliest observations of 1996. 
This excess increases in strength in 1997, and starts to dominate the 
spectrum in 1998. The stellar continuum peaks at $B$ in 1996, at $V$ 
in 1997, between $V$ and $R$ in 1998, and possibly at $R$ in 1999. From 
1998 onwards, the optical radiation of the star is extinguished and 
re-radiated at infrared wavelengths. In 1999, hardly a trace of the 
stellar contribution is visible in the energy distribution. 

A preliminary discussion of the infrared energy distributions is 
given by Kipper (1999)\markcite{kip99}. Several test runs using {\sc Dusty}
(Ivezi\'c, Nenkova \& Elizur 1997)\markcite{ive97} with parameters 
similar to those chosen by Kipper yielded non-optimal fits, which can 
possibly be explained by the different adopted value of the interstellar 
extinction and the different choice of the stellar atmosphere. A detailed 
analysis of the spectral energy distribution will be the subject of 
a future investigation.

Figure~\ref{fig8} permits estimates of the properties of the infrared 
excess, the character of the dust and the size of the dust forming 
region, and some qualitative estimates will be given. 
The radiation maximum of the infrared excess shifts towards longer 
wavelengths at later times. If the excess is approximated by 
a blackbody, its temperatures are $\sim$ 3500, 3000 $\rightarrow$ 2000, 
830 and 725 K in 1996, 1997, 1998 and 1999, respectively. The 
temperatures of 1996 and 1997 are too high for the formation of carbon dust.

The infrared excess of 1996 may be explained by free-free emission in the
wind or expanding atmosphere of V4334 Sgr, which at that time had a
surface temperature of $\sim 7500$K. In 1997, the situation is not as 
clear. Woitke, Goeres \& Sedlmayr (1996)\markcite{woi96} have shown that 
carbon nucleation can take place in shocks that occur in pulsating R CrB stars
with effective surface temperatures of 7000 K. Since V4334 Sgr 
provides similar conditions, patchy dust formation may be possible
in 1997. From 1998 onward, observational evidence of dust formation 
is beyond doubt, as was shown in Sect.~\ref{colevo}. 

\begin{figure}
\psfig{figure=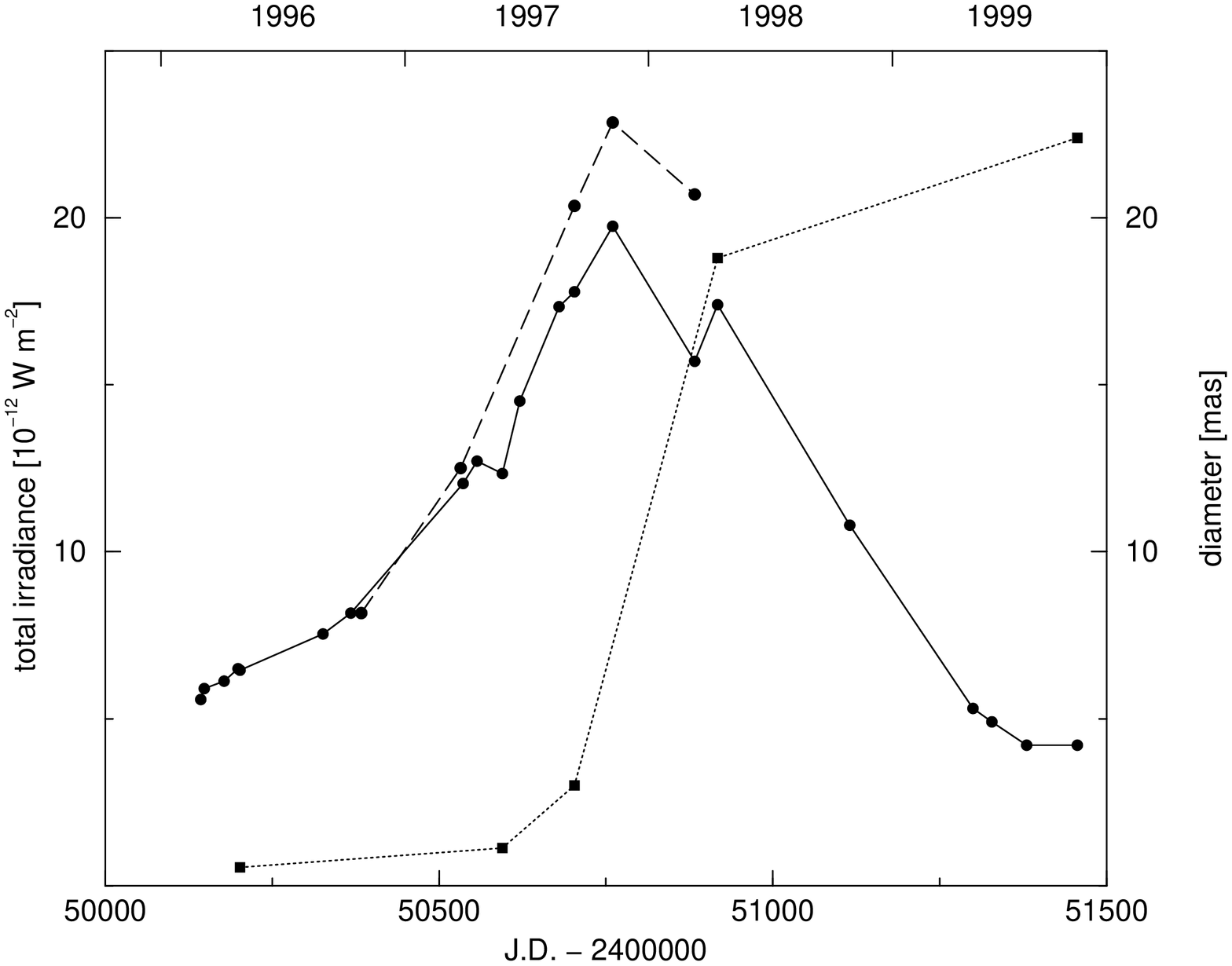,height=7cm,angle=270}
\figcaption[duerbeck.fig9.ps]{The temporal evolution of the luminosity of 
V4334 Sgr and the diameter of its radiating dust shell. The circles,
connected by solid lines, give the total irradiance shortward of 
$\rm 4.5~\mu m$; the circles, connected with long dashes, give
the total irradiance up to $\rm 14.9~\mu m$ (ISO data). The 
squares, connected by dots, give the diameter of the dust shell in
milli-arcseconds.
\label{fig9}}
\end{figure}

The angular radius of the infrared emission region, calculated from 
$\theta = 2\times 10^{12} (\lambda F_\lambda)_{\rm max}^{1/2} T^{-2}$ 
(Gallagher \& Ney 1976)\markcite{gal76} with $\theta$ in milli-arcsec 
and $(\lambda F_\lambda)_{\rm max}$ in $\rm W\,m^{-2}$, is 0.3, 0.8, 9.4 
and 11.4 milli-arcsec for the years 1996 -- 1999, respectively. 
Figure~\ref{fig9} indicates that a rapid growth of the dust shell 
occurred between late 1997 and early 1998. 

Assuming a distance of 2 kpc for the object, the radius of the dust 
shell was about 19 AU in 1998, and 23 AU in 1999. If the ejection of 
material started in early 1995, at the time of the beginning of the 
final He flash, and if dust had condensed everywhere in the shell 
in mid-1998, the constant expansion velocity $\rm 25~km\,s^{-1}$ 
of dust-forming material is derived. Between mid-1998 and mid-1999, 
the shell grew with a linear velocity of $\rm 20~km\,s^{-1}$, 
which is in good agreement with the previous value. This velocity is 
much larger than the rate of growth of the photosphere (about 
$\rm 1~km\,s^{-1}$), which was derived from the observations of 1996 
(D97). Thus we may take it as evidence for a Eddington-driven outflow 
{\it above} the photosphere, which had been active since the beginning of 
the final helium flash, and had cooled to temperatures suitable for dust 
formation around the end of 1997. 

On the other hand, we can assume that the material was ejected at a later phase
of the outburst, when the outer layers had already been enriched 
with carbon, and that the condensing dust experienced an acceleration 
due to radiation pressure. Then the resulting expansion velocity is 
higher, and may be similar to the velocity of the central condensations 
of the remnant of V605 Aql ($\rm 100~km~s^{-1}$ with a FWHM of $\rm 
225~km~s^{-1}$, Pollacco et al. 1992). More observations are necessary 
to decide between both scenarios.

\subsection{The infrared evolution of V4334 Sgr and in classical novae}
\label{evoclas}

Final He flash objects and R CrB type stars have similar properties, 
as described in Sect.~\ref{dust}. Similarities, however, also exist 
between final He flash objects and dust-forming classical novae. 

The infrared behavior of dust-forming classical novae consists of four 
phases: (a) the initial pseudo-photosphere blackbody, (b) a free-free 
phase, which leads to an infrared excess between 1 -- 6 $\mu m$, (c) 
a rapid growth of dust, leading to a ``red decline'' in the optical 
and an increase of infrared flux, of angular diameter, and of a slight 
drop in dust temperature ($\rm 1200 \rightarrow 800~K$), and (d) an 
exponential drop of the infrared flux and of the angular diameter, 
when dust is being dispersed and/or destroyed by radiation from the 
central source, accompanied by a recovery of UV and optical radiation 
(Ney \& Hatfield 1978)\markcite{ney78}. 

In 1996 and 1997, the derived blackbody temperature for the infrared 
excess of V4334 Sgr is too high for dust formation; it was already 
pointed out that the infrared excess of 1996 (and possibly 1997) 
is caused by free-free emission in the outflowing material that had 
passed the pseudo-photosphere and formed an extended atmosphere. 
Claims of the presence of dust with temperatures of 1500, 1800 and 680 K,
in 1997 February, March and April (Kerber et al. 1999\markcite{ker99},
Kimeswenger et al. 1997\markcite{kim97}, Eyres et al. 1998a\markcite{eyr98a}) 
are questionable and discrepant; part of the infrared excess may be
carbon nucleation products, part of it may still be explained by free-free 
continuum emission. 

At the end of 1997 or at the beginning of 1998, the temperature of the
extended atmosphere had dropped below 2000 K, permitting the formation 
of carbon dust. This  onset of dust formation lead to (1) a rapid growth 
of the angular size of the infrared emitting region (Fig.~\ref{fig9}), 
(2) an increase of the infrared flux, and (3) several dust forming events 
in the line of sight, which caused first a gentle, then a 
dramatic drop in visible light output (Sect.~\ref{colevo}). 

Comparing the evolution of V4334 Sgr with that of a dust-forming 
classical nova, e.g. NQ Vul (Ney \& Hatfield 1978)\markcite{ney78} 
yields the following: Phase (a) occurred likely in 1995 but was not observed, 
phase (b) occurred in 1996 -- 1997, and phase (c) in 1998 -- 1999. 
The onset of phase (d), the destruction or dispersion of the dust, 
may still be far in the future: 80 years after outburst, the central 
object of V605 Aql is still deeply embedded in circumstellar dust,
and we can expect a similar behavior for V4334 Sgr. 

Tthe ``speed'' of a classical nova like NQ Vul to evolve through phases 
(a) to (c) is about 20 times faster than that of the final He flash 
object V4334 Sgr. One more noteworthy difference between a 
nova and a final He flash object exists: 
the spectrum emerging from the pseudo-photosphere of the nova near 
maximum light clearly reveals the speed of the outflowing material, 
and the expansion rate of the infrared dust shell (in milli-arcsecond 
$\rm day^{-1}$) can be used to determine the shell expansion parallax. 
High resolution spectra of V4334 Sgr described by D97, Kipper \& Klochkova 
(1997)\markcite{kip97} and Jacoby et al. (1998)\markcite{jac98} show 
that the radial velocity of the photosphere is similar to that of the 
planetary nebula. In an echelle spectrum, however, taken 1996 April 23
by G. Wallerstein, the deep H$\alpha$ absorption line shows an underlying 
shallow, broad absorption trough ranging from $-225$ to $\rm +170~km~s^{-1}$
relative to the star. This may be taken as evidence for optically thin, 
turbulent material which shows an average outflowing velocity of $\rm 
\sim 25~km~s^{-1}$, with a wide spread in velocities. A careful study 
of spectral features of the wind, in combination with data of the growth of 
the dust shell, may permit the derivation of a shell expansion parallax. 

Dust-forming novae show optically thick winds with high outflow velocities
that exist for time scales of weeks; V4334 Sgr shows an optically thin 
wind with an outflow velocity of $\rm 25~km~s^{-1}$ (or several times higher),
which likely exists during all stages of the outburst. The increase of
luminosity at later stages (Sect.~\ref{varlum}), in combination with the 
cooling of the outer layers, may lead to a enhanced mass loss at later times.

\subsection{Variations in luminosity and limits on the mass of V4334 Sgr}
\label{varlum}

The data of Table~\ref{tab4} can be used to study the luminosity of V4334 
Sgr at various stages of ite evolution. The total irradiances (in units 
of $\rm 10^{-12}~W~m^{-2}$) are also shown in 
Fig.~\ref{fig9}. From 1997 onward, an ever increasing part of the luminosity 
is radiated at wavelengths longward of $\rm 4.5~\mu m$, and nothing 
quantitative can be said about the luminosity evolution after early 1998, 
because far infrared data are lacking.
 
The flux increased by a factor 4 from early 1996 to early 1998. 
This result depends only weakly on the assumed value of the interstellar 
extinction. It was already noted by D97 that the assumption of a constant 
luminosity of V4334 Sgr was not valid for the 1996 -- 1997 light curve; 
the flux in the optical region increased by at least 30\% over one year. 
A possible explanation of this behavior is that in early stages of the
flash, a significant fraction of the energy release is 
used for the expansion of the object (this is also shown, implicitly, 
in theoretical tracks of final He flash objects, e.g. Figs. 14 
and 15 in Bl\"ocker 1995)\markcite{blo95}. 

We take the ``late'', 1997 -- 1998 luminosity of V4334 Sgr as the 
luminosity emerging from the remnant after most of the expansional work 
had been done. A total irradiance of $2.2\times 10^{-11}~\rm W~m^{-2}~$ is 
assigned to V4334 Sgr. This can easily be converted into a radiant flux 
release of $ 10^{30} \left (\frac{d}{2~{\rm kpc}} \right)^2$~W, or
$$
L_{\rm V4334~Sgr} \sim 2770~L_\odot \left (\frac{d}{2~{\rm kpc}} \right)^2
$$

A high-mass post-AGB model of Bl\"ocker (1995)\markcite{blo95} has a 
mass of $0.836~M_\odot$, and takes 50 years for its way from the 
planetary nebula nucleus back to the AGB during the final He flash; 
its mass may be taken as a lower limit to the mass of V4334 Sgr. 
The luminosity of the model is $20,000~L_\odot$,  which may serve as a 
reasonable lower limit to the luminosity, and thus to the distance 
of V4334 Sgr. Insertion in the above relation yields 5.4 kpc for
V4334 Sgr. Even low-mass models of $\sim 0.6~M_\odot$ yield distances 
$\rm > 2~kpc$, and the assumption of Sect.~\ref{prehistory}, 
$d > 2~\rm kpc$, is always fulfilled. If one would use the range of 
absolute $B$-magnitudes of central stars of planetary nebulae as a 
way to estimate the mass of V4334 Sgr, the upper distance limit of 
3 kpc yields a mass slightly above $0.6~M_\odot$, which is somewhat 
unlikely in view of its fast evolution (see below and Sect.~\ref{time}). 
Diagrams showing evolutionary speeds, envelope and core masses for born-again 
giants, as given by Bl\"ocker \& Sch\"onberner (1997)\markcite{blo97}, 
do not cover the rapid evolution of V4334 Sgr. Nevertheless, an envelope 
mass of $10^{-5}~M_\odot$ and a core mass around $1~M_\odot$ are 
reasonable guesses. The high luminosity of such an object may even 
give support to the ``long'' distance scale of 8 kpc as suggested 
by D97. Further research on post-AGB evolution is clearly needed to
constrain the distance of V4334 Sgr. 
 
\section{Time scales}
\label{time} 

Data on the evolution of two previous final He flash objects exist: FG Sge 
and V605 Aql. We omit from our discussion the 17th century object CK Vul, 
whose nature is still not clear and whose light curve covers 
only the brightest stages (Harrison 1996\markcite{har96}). Data were 
taken from Harrison (1996)\markcite{har96}, Clayton \& De Marco 
(1997)\markcite{cla97}, and Jurcsik \& Montesinos (1999)\markcite{jur99}, 
and compared with the present data. Table~\ref{tab5} gives the time 
scales involved.

\placetable{tab5}

The only existing high-quality spectrum of V605 Aql was described by 
Bidelman (1973)\markcite{bid73} to be ``very similar to the hydrogen-deficient
carbon star HD 182040'', which has a type C2,2 in the old Keenan-Morgan 
classification, and C-HD1$\rm C_24^-$CH0 in the 1993 revised MK system 
of Keenan (Barnbaum, Stone \& Keenan 1996)\markcite{bar96}. The appearance 
of the spectrum of V4334 Sgr, taken in May 1997, and analyzed by Pavlenko, 
Yakovina \& Duerbeck (2000)\markcite{pav00}, is strikingly similar to that 
of V605 Aql, as illustrated by Clayton \& De Marco (1997)\markcite{cla97}, 
and classified by Bidelman.

We compare the light curves of  FG Sge, V605 Aql and V4334 Sgr in 
detail, taking the light curve of Harrison (1996)\markcite{har96} 
for V605 Aql, and assuming that maximum $B$ (or photographic) light 
occurred in 1968, 1919.6 and 1996.3, for the three objects, 
respectively. Note that Harrison's light curve of V605 Aql shows a 
minimum already in 1920; this ``first'' dust event, however, seems 
to be poorly documented and will not be taken into consideration here.
The following time intervals are derived: Rise from about $15^{\rm m}$
to ($B$ or photographic) maximum, took 74, 1.9 and 1.5 years, respectively. 
A comparable spectral type C2,2 was reached about 20 years, 2.1 and 
1.0 years after maximum. Dust event onset, first dust event and 
first minor dust event were observed
24, 3 and 2.1 years after maximum. ``Disappearance'' due to a major dust 
event has not yet been observed for FG Sge, and occurred 4.4 and 2.9 
years after maximum for the two other objects. The ``total duration 
of visibility'' (with moderate means) is thus 6.5 and 4.4 years for 
V605 Aql and V4334 Sgr, respectively. FG Sge has not yet entered the 
stage of faintness, and one can only compare the time from the onset 
of brightening to the onset of dust formation, which is 98 years for FG
Sge, 4.9 years for V605 Aql, and 3.6 years for V4334 Sgr. ``Averaging'' 
the timescales of various events in the three objects, one finds that 
V4334 Sgr is the most rapidly evolving (and presumably the most massive) 
final He flash object known; V605 Aql is about 50\% slower, and FG Sge 
is a factor of 25 -- 50 slower. Extrapolating the lifetime of FG Sge 
to its expected disappearance due to a future major dust event yields 
a value of up to 220 years. Already one half of this time has 
elapsed; it will be interesting to monitor the future evolution
of FG Sge.

\section{Summary and outlook}
\label{summary}

The complete multi-color light curve of V4334 Sgr from its pre-discovery
rise to the dust obscuration shows that the color indices increase
quite smoothly. In 1995 -- 1997, this is caused by 
the cooling of the expanding pseudo-photosphere of a mass-losing object 
that has a slowly increasing luminosity. Furthermore, the increasing
infrared excess can be explained by free-free emission in an 
Eddington-driven outflow. Starting from 1998, brightness drops and 
their color characteristics mimic the ``red declines'' of R CrB 
variables. The increase in infrared flux and the behavior of the 
collisionally excited \ion{He}{1} 10830 line indicate that a complete 
dust shell formed around the object in late 1998 -- early 1999.
Such a phenomenon also occurs in dust-forming classical novae. 

The dust formation in V4334 Sgr, and possibly in most massive 
final He flash objects is ``catastrophic'', i.e. a shell is formed which 
surrounds the whole star and which does not dissipate quickly. V605 Aql, 
after its disappearance in 1924, never recovered from its dust episode: 
plates of the Sonneberg sky patrol from 1928 -- 1979, reaching (mostly 
photographic) magnitude 16 -- 17.5, did not recover it 
(Fuhrmann 1981)\markcite{fuh81}. It was only recovered at a very faint
magnitude (Seitter 1985). It is quite certain that V4334 Sgr will 
behave in a similar way in the years to come.

Thus, final He flash objects show some similarity to R CrB stars, 
but apparently the onset of R CrB-like activity, at least for the 
quickly evolving objects like V605 Aql and V4334 Sgr, soon ends in 
a ``catastrophic'' decline, and does not extend over centuries of 
stellar evolution. The slowly evolving object FG Sge is also much 
more active than normal R CrB stars, but it has shown ``blue 
declines'', indicating that only localized dust formation has 
occurred until now. It will be extremely interesting to follow 
the future behavior of FG Sge. Final He flash objects as we know 
them are obviously not settling down as ``normal'' R CrB stars. 
Possibly low-mass, very slowly evolving final He flash objects 
(with FG Sge possibly defining the high mass limit) are the ones 
that may show up as R CrB stars during extended evolutionary phases.

The rapid evolution of V4334 Sgr (and of V605 Aql) indicate that we
observe here the massive objects undergoing post-AGB evolution.
Models covering such masses and timescales are badly needed, in order to
constrain masses, luminosities and distances of the observed events.

\acknowledgements
 
This paper profited much from a stay of H.W.D. at STScI Baltimore. 
He thanks M. Shara and N. Panagia for support and
encouragement, and K. Sahu for arranging a seminar talk. 
Helpful electronic discussions with M. Asplund (Uppsala),
A. Evans (Keele), U.S. Kamath and N.M. Ashok (Ahmedabad) and
Ya. Pavlenko (Kiev) are gratefully recognized, and we are also very much
indebted to P.A. Whitelock (SAAO) for communicating infrared data in advance
of publication, to G. Wallerstein (Seattle) for communicating 
spectroscopic observations, and to W.C. Seitter for a careful reading of the 
manuscript. C.S. and T.A. acknowledge
financial support from the Fund for Scientific Research Flanders (FWO).
This research was supported by the Belgian Fund for Scientific Research (FWO)
and by the Flemish Ministry for Foreign Policy, European Affairs, Science and
Technology. Finally, we acknowledge the comments of a referee that were
very helpful in improving the presentation of the paper.

\begin{table}
\tablenum{1}

\caption{{\it UBVRi} observations of V4334 Sgr\label{tab1}}

\begin{tabular}{lrrrrrc}\hline
J.D.hel.  & $U$     & $B$    & $V$    & $R_C$  & $i_G$  & observer \\ 
2450000+  &         &        &        &        &        &          \\ \hline
504.884 &         &        & 10.95  &        &        & WL \\
514.867 &         &        & 11.05  &        &  9.18  & TV \\
515.877 &         &        & 11.04  &        &  9.18  & TV \\
516.889 &         &        & 11.04  &        &  9.18  & TV \\
516.913 &         &        & 11.04  &        &        & WL \\
517.882 &         &        & 11.04  &        &  9.19  & TV \\
518.875 &         &        & 11.04  &        &  9.19  & TV \\
519.889 &         &        & 11.04  &        &  9.20  & TV \\ 
520.901 &         &        & 11.02  &        &  9.20  & TV \\
521.907 &         &        & 11.02  &        &  9.18  & TV \\
521.906 &         &        & 10.99  &        &        & WL \\
522.884 &         &        & 11.02  & 10.01  &  9.18  & TV \\
523.861 &         &        & 11.02  &        &  9.18  & TV \\
539.914 &  13.49  & 12.60  & 10.92  &  9.93  &  8.96  & MT \\
555.896 &  13.32  & 12.46  & 10.86  &  9.82  &  8.94  & SB \\
556.924 &  13.37  & 12.50  & 10.86  &  9.86  &  8.94  & SB \\
557.728 &         &        & 10.82  &        &        & WL \\ 
563.769 &         &        & 10.95  &        &        & WL \\
570.701 &         &        & 11.10  &        &        & WL \\
574.710 &         &        & 11.12  &        &        & WL \\
581.679 &         &        & 11.06  &        &        & WL \\
587.803 &         &        & 11.09  &        &        & WL \\
590.705 &         &        & 11.03  &        &        & WL \\
593.660 &  13.40  & 12.61  & 11.00  &  9.93  &  9.04  & EB \\
595.795 &  13.47  & 12.64  & 10.97  &  9.92  &  9.02  & EB \\
596.723 &  13.49  & 12.60  & 10.96  &  9.91  &  9.04  & EB \\
600.822 &         & 12.57  & 10.96  &  9.91  &  8.98  & EB \\
601.608 &  13.51  & 12.62  & 10.96  &  9.90  &  8.98  & EB \\
601.640 &         &        & 10.94  &        &        & WL \\
604.577 &         &        & 10.93  &        &        & WL \\
605.771 &  13.53  & 12.61  & 10.96  &  9.87  &  8.97  & EB \\
605.644 &         &        & 10.93  &        &        & WL \\
606.857 &  13.41  & 12.58  & 10.95  &  9.86  &  8.94  & EB \\
607.670 &  13.43  & 12.57  & 10.94  &  9.84  &  8.94  & EB \\
608.824 &  13.48  & 12.57  & 10.97  &  9.86  &  8.94  & EB \\
613.763 &  13.55  & 12.58  & 10.98  &  9.89  &  8.95  & EB \\
615.630 &  13.44  & 12.61  & 10.98  &  9.88  &  8.95  & EB \\
615.656 &         &        & 10.97  &        &        & WL \\
616.840 &  13.44  & 12.61  & 10.97  &  9.86  &  8.94  & EB \\ 
617.614 &         &        & 10.93  &        &        & WL \\
618.712 &  13.46  & 12.60  & 10.97  &  9.88  &  8.94  & EB \\
621.648 &         &        & 10.89  &        &        & WL \\
623.748 &         &        & 10.96  &        &        & WL \\
624.506 &         &        & 10.96  &        &        & WL \\
625.496 &         &        & 10.93  &        &        & WL \\
626.501 &         &        & 10.93  &        &        & WL \\
627.486 &         &        & 10.88  &        &        & WL \\
628.510 &         &        & 10.93  &        &        & WL \\
629.539 &         &        & 10.96  &        &        & WL \\
630.503 &         &        & 10.93  &        &        & WL \\
\end{tabular}
\end{table}

\begin{table}
\begin{tabular}{lrrrrrc}\hline
J.D.hel.  & $U$     & $B$    & $V$    & $R_C$  & $i_G$  & observer \\ 
2450000+  &         &        &        &        &        &   \\ \hline
632.515 &         &        & 10.89  &        &        & WL \\
633.546 &         &        & 10.91  &        &        & WL \\
634.499 &         &        & 10.88  &        &        & WL \\
640.569 &         &        & 10.97  &        &        & WL \\
642.503 &         &        & 10.91  &        &        & WL \\
643.482 &         &        & 10.92  &        &        & WL \\
644.482 &         &        & 10.91  &        &        & WL \\
645.478 &         &        & 10.92  &        &        & WL \\
646.474 &         &        & 10.96  &        &        & WL \\
649.478 &         &        & 10.94  &        &        & WL \\
651.516 &         &        & 10.86  &        &        & WL \\
652.500 &         &        & 10.86  &        &        & WL \\
653.497 &         &        & 10.89  &        &        & WL \\
655.512 &         &        & 10.87  &        &        & WL \\
661.476 &         &        & 10.85  &        &        & WL \\
662.484 &         &        & 10.77  &        &        & WL \\
683.548 &  13.38  & 12.50  & 10.85  &  9.74  &  8.77  & JK \\
684.542 &  13.45  & 12.49  & 10.86  &  9.74  &  8.79  & JK \\
685.506 &         &        & 10.86  &        &  8.80  & JK \\
686.502 &         &        & 10.84  &        &  8.80  & JK \\
687.520 &         &        & 10.86  &        &  8.82  & JK \\
687.597 &         &        & 10.87  &        &  8.82  & JK \\
689.496 &         &        & 10.87  &        &  8.82  & JK \\
690.509 &         &        & 10.88  &        &  8.82  & JK \\
692.468 &         &        & 10.92  &        &  8.83  & JK \\
694.496 &  13.50  & 12.58  & 10.92  &  9.80  &  8.87  & JK \\
696.506 &         &        & 10.94  &        &  8.86  & JK \\
697.493 &         &        & 10.95  &        &  8.86  & JK \\
698.495 &         &        & 10.97  &        &  8.87  & JK \\
699.490 &         &        & 10.96  &        &  8.87  & JK \\
700.487 &         &        & 10.96  &        &  8.86  & JK \\
701.488 &         &        & 10.95  &        &  8.86  & JK \\
703.490 &         &        & 10.94  &        &  8.85  & JK \\
704.491 &         &        & 10.94  &        &  8.86  & JK \\
705.508 &  13.40  & 12.60  & 10.94  &  9.83  &  8.86  & JK \\
706.502 &         &        & 10.93  &        &  8.86  & JK \\
707.498 &  13.44  & 12.60  & 10.93  &  9.81  &  8.86  & JK \\
712.468 &         & 12.56  & 10.88  &        &        & CS \\
713.463 &         & 12.54  & 10.85  &        &        & CS \\
714.471 &         & 12.53  & 10.85  &        &        & CS \\
722.535 &         &        & 10.79  &        &        & WL \\
725.486 &         &        & 10.86  &        &        & WL \\
727.522 &         &        & 10.80  &        &        & WL \\
728.494 &         &        & 10.80  &        &        & WL \\
737.477 &         & 12.56  & 10.87  &        &        & CS \\
741.485 &         & 12.58  & 10.90  &        &        & CS \\
743.500 &         &        & 10.86  &        &        & WL \\
743.506 &         &        & 10.92  &        &        & CS \\
745.517 &         &        & 10.92  &        &        & WL \\
746.496 &         &        & 10.93  &        &        & WL \\
\end{tabular}
\end{table}

\begin{table}
\begin{tabular}{lrrrrrc}\hline
J.D.hel.  & $U$     & $B$    & $V$    & $R_C$  & $i_G$  & observer \\ 
2450000+  &         &        &        &        &        & \\ \hline
750.511 &         &        & 10.90  &        &        & WL \\
752.499 &         &        & 10.90  &        &        & WL \\
761.513 &         &        & 10.88  &        &        & WL \\
765.519 &         &        & 10.84  &        &        & WL \\
769.520 &         &        & 10.75  &        &        & WL \\
773.522 &         &        & 10.76  &        &        & WL \\
853.850 &         &        & 11.88  &        &        & WL \\
853.886 &         &        & 12.20  &        &        & WL \\
854.853 &         &        & 12.15  &        &        & WL \\
856.885 &         &        & 12.22  &        &  9.78  & MJ \\
857.900 &  15.21  & 14.17  & 12.27  & 10.97  &  9.83  & MJ \\
858.842 &  14.96  & 14.17  & 12.30  & 10.98  &  9.84  & MJ \\
859.882 &  15.04  & 14.24  & 12.34  & 11.03  &  9.87  & MJ \\
860.877 &  15.20  & 14.27  & 12.36  & 11.04  &  9.88  & MJ \\
862.858 &  15.10  & 14.26  & 12.34  & 11.05  &  9.90  & MJ \\
865.899 &         & 14.32  & 12.36  & 11.05  &  9.85  & MJ \\
866.864 &         & 99.00  & 12.34  & 11.05  &  9.86  & MJ \\
868.882 &  15.31  & 14.31  & 12.36  & 11.04  &  9.89  & MJ \\
869.894 &  15.22  & 14.35  & 12.37  & 11.06  &  9.89  & MJ \\
870.399 &  15.24  & 14.37  & 12.40  & 11.07  &  9.89  & MJ \\
871.894 &  15.16  & 14.46  & 12.40  & 11.08  &  9.90  & MJ \\
872.393 &  15.16  & 14.40  & 12.41  & 11.09  &  9.91  & MJ \\
873.893 &  15.37  & 14.41  & 12.41  & 11.10  &  9.92  & MJ \\
874.897 &  15.28  & 14.45  & 12.43  & 11.11  &  9.91  & MJ \\
875.841 &         &        & 12.40  &        &        & WL \\
875.899 &  15.36  & 14.47  & 12.43  & 11.11  &  9.92  & MJ \\
876.901 &  15.48  & 14.48  & 12.46  & 11.11  &  9.92  & MJ \\
877.909 &  15.42  & 14.46  & 12.46  & 11.10  &  9.92  & MJ \\
878.906 &  15.47  & 14.48  & 12.45  & 11.11  &  9.91  & MJ \\
879.883 &         &        & 12.54  &        &        & WL \\
880.876 &         &        & 12.55  &        &        & WL \\
881.911 &  15.44  & 14.48  & 12.43  & 11.08  &  9.89  & MJ \\
887.847 &         & 14.33  & 12.30  &        &        & CS \\
888.843 &         & 14.31  & 12.29  &        &        & CS \\
890.825 &         &        & 12.33  &        &        & WL \\
890.869 &         & 14.28  & 12.17  &        &        & CS \\
891.888 &         &        & 12.34  &        &        & WL \\
891.896 &         & 14.26  & 12.23  &        &        & CS \\
892.811 &         & 14.25  & 12.22  &        &        & CS \\
893.915 &  15.52  & 14.28  & 12.22  & 10.86  &  9.71  & HD \\
894.850 &         &        & 12.34  &        &        & WL \\
894.921 &  15.38  & 14.22  & 12.20  & 10.85  &  9.68  & HD \\
895.885 &  15.31  & 14.23  & 12.20  & 10.85  &  9.68  & HD \\
896.809 &         &        & 12.29  &        &        & WL \\
897.816 &         &        & 12.36  &        &        & WL \\
899.837 &         &        & 12.20  &        &        & WL \\
901.818 &         &        & 12.22  &        &        & WL \\
909.872 &         &        & 12.22  &        &        & WL \\
910.788 &         &        & 12.25  &        &        & WL \\
918.881 &         & 14.02  & 12.01  &        &        & TA \\
\end{tabular}
\end{table}

\begin{table}
\begin{tabular}{lrrrrrc}\hline
J.D.hel.  & $U$     & $B$    & $V$    & $R_C$  & $i_G$  & observer \\ 
2450000+  &         &        &        &        &        & \\ \hline
919.871 &         &        & 12.11  &        &        & WL \\
921.753 &         &        & 12.19  &        &        & WL \\
922.793 &         &        & 11.97  &        &        & TA \\
923.693 &         &        & 12.14  &        &        & WL \\
925.670 &         &        & 12.19  &        &        & WL \\
926.782 &         &        & 12.11  &        &        & WL \\
927.701 &         &        & 12.02  &        &        & WL \\
931.770 &         &        & 11.93  &        &        & WL \\
934.764 &         &        & 11.84  &        &        & WL \\
937.716 &         &        & 11.79  &        &        & WL \\
941.741 &         &        & 11.84  &        &        & WL \\
949.801 &         &        & 11.88  &        &        & WL \\
950.723 &         &        & 11.65  &        &  9.28  & PW \\
963.625 &         &        & 12.20  &        &        & WL \\
965.711 &         &        & 12.39  &        &        & WL \\
971.701 &         &        & 12.36  &        &        & WL \\
980.628 &         &        & 12.07  &        &        & WL \\
994.498 &         &        & 11.86  &        &        & WL \\
996.647 &         &        & 11.77  &        &        & WL \\
997.508 &         &        & 11.78  &        &        & WL \\
1000.505 &         &        & 11.71  &        &        & WL \\
1004.522 &         &        & 11.66  &        &        & WL \\
1007.495 &         &        & 11.65  &        &        & WL \\
1011.517 &         &        & 11.65  &        &        & WL \\
1012.478 &         &        & 11.66  &        &        & WL \\
1013.542 &         &        & 11.71  &        &        & WL \\
1014.466 &         &        & 11.67  &        &        & WL \\
1016.520 &         &        & 11.72  &        &        & WL \\
1019.506 &         &        & 11.84  &        &        & WL \\
1020.491 &         &        & 11.77  &        &        & WL \\
1023.484 &         &        & 12.02  &        &        & WL \\
1026.517 &         &        & 12.06  &        &        & WL \\
1029.486 &         &        & 12.27  &        &        & WL \\
1031.489 &         &        & 12.12  &        &        & WL \\
1034.510 &         &        & 12.36  &        &        & WL \\
1034.717 &  15.58  & 14.26  & 12.29  & 10.95  &  9.80  & JA \\
1035.531 &         &        & 12.33  &        &  9.80  & JA \\
1036.481 &         &        & 12.48  &        &        & WL \\
1036.773 &         &        & 12.38  &        &  9.86  & JA \\
1037.520 &         &        & 12.30  &        &        & WL \\
1037.640 &         &        & 12.47  &        &  9.94  & JA \\
1038.527 &         &        & 12.55  &        &        & WL \\
1038.664 &         &        & 12.54  &        &  9.99  & JA \\
1039.492 &         &        & 12.68  &        &        & WL \\
1039.637 &         &        & 12.61  &        & 10.04  & JA \\
1040.752 &         &        & 12.63  &        & 10.07  & JA \\
1041.610 &  16.26  & 14.81  & 12.72  & 11.34  & 10.15  & JA \\
1042.700 &         &        & 12.76  &        & 10.15  & JA \\
1043.519 &         &        & 12.76  &        &        & WL \\ 
1043.549 &         &        & 12.80  &        & 10.19  & JA \\
\end{tabular}
\end{table}

\begin{table}
\begin{tabular}{lrrrrrc}\hline
J.D.hel.  & $U$     & $B$    & $V$    & $R_C$  & $i_G$  & observer \\ 
2450000+  &         &        &        &        &        & \\ \hline
1044.696 &         &        & 12.84  &        & 10.23  & JA \\
1045.677 &         &        & 12.88  &        & 10.26  & JA \\
1046.666 &         &        & 12.92  &        & 10.28  & JA \\
1047.647 &         &        & 12.94  &        & 10.30  & JA \\
1048.664 &  16.74  & 15.11  & 12.98  & 11.54  & 10.30  & JA \\
1049.549 &         &        & 12.98  &        & 10.32  & JA \\
1051.624 &  16.82  & 15.18  & 13.01  & 11.56  & 10.35  & JA \\
1052.724 &         &        & 12.98  &        & 10.36  & JA \\
1053.481 &         &        & 13.04  &        & 10.38  & JA \\
1054.541 &         &        & 13.07  &        & 10.38  & JA \\
1055.549 &  16.84  & 15.22  & 13.07  & 11.64  & 10.40  & JA \\
1059.668 &         & 15.28  & 13.14  & 11.68  & 10.46  & JA \\
1061.659 &         &        & 13.20  &        & 10.55  & JA \\
1062.661 &  17.01  & 15.34  & 13.27  & 11.82  & 10.59  & JA \\
1063.576 &  17.11  & 15.53  & 13.32  & 11.88  & 10.63  & CS \\
1064.579 &         & 15.50  & 13.39  &        &        & CS \\
1065.578 &  17.22  & 15.55  & 13.43  & 11.97  & 10.74  & CS \\
1067.593 &  17.40  & 15.74  & 13.64  & 12.16  & 10.91  & CS \\
1068.626 &  17.56  & 15.85  & 13.72  & 12.25  & 10.98  & CS \\
1069.619 &  17.66  & 15.99  & 13.85  & 12.38  & 11.08  & CS \\
1070.554 &  17.83  & 16.11  & 13.96  & 12.47  & 11.19  & CS \\
1071.501 &         &        & 14.03  &        &        & WL \\
1071.625 &  17.98  & 16.24  & 14.10  & 12.55  & 11.29  & CS \\
1072.617 &  18.28  & 16.45  & 14.26  & 12.73  & 11.42  & CS \\
1074.475 &         &        & 14.23  &        &        & WL \\
1075.506 &         &        & [15.   &        &        & WL \\
1078.515 &         &        & [15.   &        &        & WL \\
1081.532 &         &        & [15.   &        &        & WL \\
1094.529 &         &        & [15.   &        &        & WL \\
1097.521 &         &        & [15.   &        &        & WL \\
1101.509 &         &        & [15.   &        &        & WL \\
1110.509 &         &        & [15.   &        &        & WL \\
1115.529 &         &        & [15.   &        &        & WL \\
1125.516 &  19.75  & 17.62  & 15.14  & 13.42  & 11.94  & AM \\
1126.524 &  19.82  & 17.60  & 15.14  & 13.43  & 11.93  & AM \\
1127.507 &         & 17.69  & 15.17  & 13.44  & 11.97  & AM \\
1128.514 &         &        & 15.20  &        & 11.98  & AM \\
1129.507 &         & 17.69  & 15.21  & 13.45  & 11.98  & AM \\
1130.503 &         &        & 15.22  &        & 12.00  & AM \\
1130.628 &         &        & 15.13  &        &        & WL \\
1234.87  &         & 19.187 & 16.45  & 14.54  & 12.82  & RD \\
1235.86  &         &        & 16.61  &        & 12.93  & RD \\
1236.89  &         &        & 16.81  &        & 13.06  & RD \\
1238.87  &         & 20.31  & 17.26  & 15.15  & 13.34  & RD \\
1260.837 &         &        & 20.10  &        &        & CS \\
1261.871 &         &        & 19.90  &        &        & CS \\
1262.857 &         &        & 19.67  &        &        & CS \\
1263.866 &         &        & 20.15  &        &        & CS \\
1264.876 &         &        & 19.92  &        &        & CS \\
1265.910 &         &        & 19.84  &        &        & CS \\
1267.865 &         &        & 20.08  & 17.72  & 15.56  & CS \\
1368.06  & [23.48  &        & 22.10  & 20.87  & 18.51  & TN \\ \tableline
\end{tabular}

\tablecomments{Observer (at 0.91 m Dutch telescope, unless stated 
otherwise): \\
JA: J.~Arts, TA: Torben Arentoft, SB: S.~Benetti, TN: S.~Benetti, 
3.5 m TNG La Palma, EB: E.~Brogt, RD: R.~Dijkstra, HD: H.W.~Duerbeck,
MJ: M.~Janson, JK: J.~Kurk,  WL: W.~Liller (0.2 m telescope, 
Re$\rm \tilde{n}$aca Bajo), AM: A.~van der Meer, CS: C.~Sterken, 
MT: M.~Turatto, TV: T.~Voskes,  PW: P. de Wildt.}

\end{table}

\begin{table}
\tablenum{2}

\caption{Absolute $B$ magnitudes of central stars of 
selected planetary nebulae\label{tab2}}

\begin{tabular}{lcccccl}\tableline
Object   & $d$ (pc)  & source & $m_B$ & $A_B$ & $M_B$ & comment\\ \tableline
NGC 6720 & 704 & USNO & 15.3\phn      & 0.21 & 5.85 & Ring nebula\\
NGC 6853 & 380 & USNO & 13.66  & 0.50 & 5.36 &\\
NGC 7293 & 213 & USNO & 13.10  & 0.08 & 6.38 & Helix nebula\\
Abell 31 & 211 & USNO & 15.20  & 0.07 & 8.51 &\\
Abell 36 & 243 & Hipp & 11.28  & 0.10 & 4.25 &\\
Abell 74 & 752 & USNO & 16.91  & 0.25 & 7.28 &\\
PHL 932  & 110 & Hipp & 11.83  & 0.09 & 6.53 &\\
PW1      & 433 & USNO & 15.3\phn      & 0.29 & 6.83 &\\
Sh 216   & 130 & USNO & \phm{pg\,}12.3\,pg\phn    & 0.26 & 6.47 &\\ \tableline
\end{tabular}

\tablecomments{USNO = parallax from US Naval Observatory; Hipp = 
parallax from Hipparcos catalogue (Jacoby, de Marco \& Sawyer 1998, 
Acker et al. 1998).} 
\end{table}

\begin{table}
\tablenum{3}

\caption{Color evolution of V4334 Sgr\label{tab3}}

\begin{tabular}{ccccccl}\tableline
J.D.  &   $V$     &  $U-B$ &  $B-V$ &  $V-R$ &  $V-i$ & reference\\ \tableline
2450136 &  11.29  &        &  0.84  &  0.53  &  1.11 & \\
2450144 &  11.22  &  0.23  &  0.87  &  0.56  &  1.15 & \\
2450173 &  11.15  &  0.28  &  0.89  &  0.58  &  1.17 & \\
2450196 &  11.06  &  0.38  &  0.92  &  0.63  &  1.18 & \\
2450287 &  11.03  &  0.58  &  1.12  &  0.68  &  1.34 & \\
2450328 &  10.91  &  0.64  &  1.19  &  0.74  &  1.42 & \\
2450350 &  10.86  &  0.67  &  1.24  &  0.76  &  1.45 & \\
2450550 &  10.88  &  0.87  &  1.64  &  1.01  &  1.93 & \\
2450595 &  10.98  &  0.83  &  1.64  &  1.06  &  1.95 & \\
2450611 &  10.96  &  0.88  &  1.62  &  1.10  &  2.02 & \\
2450684 &  10.85  &  0.92  &  1.65  &  1.12  &  2.07 & \\
2450706 &  10.94  &  0.83  &  1.66  &  1.11  &  2.08 & \\
2450895 &  12.21  &  1.16  &  2.04  &  1.36  &  2.52 & \\
2451042 &  12.66  &  1.47  &  2.06  &  1.39  &  2.58 & \\
2451059 &  13.22  &  1.64  &  2.14  &  1.45  &  2.68 & \\
2451070 &  13.92  &  1.72  &  2.14  &  1.50  &  2.78 & \\
2451094 &  18.09  &        &  2.89  &  2.31  &  4.32 & JDM \\ 
2451105 &  16.91  &        &  3.05  &  2.11  &  3.77 & JDM \\  
2451126 &  15.14  &  2.18  &  2.47  &  1.71  &  3.21 & \\
2451237 &  16.85  &        &  2.89  &  2.01  &  3.77 & \\ 
2451259 &  20.08  &        &  3.17  &  2.75  &  4.99 & J \\ 
2451267 &  20.08  &        &        &  2.36  &  4.52 & \\ 
2451368 &  22.10  &        &        &  1.23  &  3.59 & TNG \\ \tableline 
\end{tabular}

\tablecomments{JDM = Jacoby \& De Marco (1998), J = Jacoby (1999), TGN = 
Telescopio Nazionale Galilei (this paper)}
\end{table}

\begin{table}
\tablenum{4}
\setlength{\tabcolsep}{1.mm}

\caption{Dereddened monochromatic irradiances, 
in $[10^{-12}~\rm W~m^{-2}~\mu m^{-1}]$, 
of V4334 Sgr as a function of time\label{tab4}}

\tiny

\begin{tabular}{c|ccccccccccccccccc|cc}\tableline
J.D. & $U$ & $B$ & $V$ & $R$ & $I$ & $J$ & $H$ & $K$ & $L$ & LW1 &
$M$ & LW4 & LW5 & LW6 & LW7 & LW8 & LW9 & flux 
& flux \\
 & 0.36 & 0.44 & 0.55 & 0.67 & 0.81 & 1.25 & 1.62 & 2.2  & 3.5 & 4.5  & 
5.0  & 6.0  & 6.8  & 7.7  & 9.6  & 11.3 & 14.9  & ($\rm \le 4.5~\mu m$) & 
($\rm \ge 4.5~\mu m$) \\ \tableline
2450143 & 16.95 & 18.63 & 11.70 & 4.36 & 2.830 & 1.065 & 0.434 & 0.144 
& 0.025 &     &    &      &     &     &     &      &     & 5.57 & \\
2450148 & 17.75 & 19.69 & 12.59 & 4.70 & 2.963 & 1.126 & 0.459 & 0.151 
& 0.027 &     &    &      &     &     &     &      &     & 5.90 & \\
2450178 & 16.65 & 19.51 & 12.59 & 4.83 & 3.074 & 1.280 &       & 0.158 
&       &     &    &      &     &     &     &      &     & 6.13 & \\
2450199 & 17.11 & 21.19 & 13.93 & 5.29 & 3.402 & 1.353 & 0.557 & 0.184 
&       &     &    &      &     &     &     &      &     & 6.49 & \\
2450202 & 16.80 & 21.00 & 13.68 & 5.29 & 3.340 & 1.353 & 0.557 & 0.186 
& 0.035 &     &    &      &     &     &     &      &     & 6.45 & \\
2450326 & 11.62 & 18.80 & 15.56 & 6.98 & 4.784 & 2.145 & 0.925 & 0.315 
& 0.059 &     &    &      &     &     &     &      &     & 7.54 & \\
2450368 & 11.31 & 18.97 & 16.59 & 7.44 & 5.294 & 2.485 & 1.091 & 0.389 
& 0.079 &     &    &      &     &     &     &      &      &  8.16 & \\
2450536 &  5.99 & 12.08 & 15.85 & 8.79 & 7.939 & 4.240 &       & 1.592 
&       &0.21 &    & 0.10 &0.068&0.044&0.020&0.0097&0.0033& 12.05 & 0.46 \\
2450557 &  6.69 & 13.25 & 16.75 & 9.37 & 8.086 & 4.780 & 3.236 & 1.811 
& 0.813 &     &    &      &     &     &     &      &      & 12.72 & \\
2450595 &  6.51 & 11.97 & 14.72 & 8.79 & 7.375 & 4.959 & 3.266 & 1.745 
&       &     &    &      &     &     &     &      &      & 12.35 & \\
2450621 &  6.16 & 12.08 & 15.14 & 9.20 & 8.086 & 5.538 & 4.111 & 2.477 
& 1.300 &     &0.356&     &     &     &     &      &      & 14.51 & \\
2450680 &  6.63 & 13.25 & 16.91 &10.47 & 9.457 & 7.505 & 5.081 & 3.295 & 
1.845 &       &0.534&     &     &     &     &      &      & 17.34 & \\
2450703 &  6.51 & 12.08 & 15.56 & 9.63 & 8.705 & 6.720 & 5.419 & 3.419 
& 2.178 &1.04 &0.636& 0.62&0.444&0.297&0.133&0.073 & 0.026& 17.79 & 2.57 \\
2450760 &       &       & 16.44 &      &       & 7.575 & 5.888 & 3.714 
& 2.218 &1.29 &     & 0.66&0.482&0.330&0.145&0.080 & 0.028& 19.75 & 3.11 \\
2450883 & 0.994 & 2.14 & 3.94 & 3.05 & 3.371 & 4.823 & 5.224 & 4.425  
& 3.034 &1.82 &      & 1.07&0.835&0.550&0.260&0.138 & 0.047& 15.71 & 4.99 \\
2450917 &       & 3.27 & 5.81 &       &       & 5.589 & 5.834 & 4.465 
& 3.006 &     &      &     &     &     &     &      &      & 17.4: & \\
2451115 & 0.019 & 0.120 & 0.325 & 0.353 & 0.510 & 0.710 & 2.042 & 3.205 
& 3.716 &     &      &     &     &     &     &      &      & 10.8: & \\
2451300 &       &       & 0.004 &       &       & 0.079 & 0.589 & 1.506 
& 2.900 &     &      &     &     &     &     &      &      &  5.3: & \\
2451328 &       &       &       &       &       & 0.069 & 0.490 & 1.336 
& 2.643 &     &      &     &     &     &      &      &      & 4.9: & \\
2451380 &       &     & 0.0005  & 0.0004 & 0.0012 & 0.019 & 0.250 & 0.859 
&       &     &      &     &     &     &      &      &      & 4.2: & \\
2451456 &       &       &       &       &       &       & 0.232 & 0.908 
& 2.667 & 2.0 &      &     &     &     &      &      &      & 4.2: 
& \\ \tableline
\end{tabular}

\end{table}

\begin{table}
\tablenum{5}

\caption{Events in the evolution of FG Sge, V605 Aql and V4334 Sgr\label{tab5}}

\begin{tabular}{l|lll}\tableline
Object  & FG Sge  & V605 Aql & V4334 Sge \\ \tableline
brightness increase (spectrum) & 
1894-1975 [B-G2 I] & 1917.7-1918 & 1994.8-1995 \\
time of brightness maximum in $B$ (spectrum) & 
1968 [A3 I]     & 1919.6    & B 1996.3 [F0] \\
spectrum at later stage &  
G-K0 I in 1980s & C2,2 in 1921.7 & C2,2 in 1997.3\\
onset of dust formation &
1992            & 1922.6 (?) &  1998.4 \\
dramatic decline (``disappearance'') &
 ?                       & 1924   & 1999.2 \\ \tableline
\end{tabular}
\end{table}

\begin{references}
\reference{ack92}Acker, A., Ochsenbein, F., Stenholm, B.,
Tylenda, R., Marcout, J., Schohn, C. 1992, Strasbourg-ESO Catalogue of
Galactic Planetary Nebulae, Garching: European Southern Observatory
\reference{ack98}Acker, A., Fresneau, A., Pottasch, S.R., Jasniewicz, G. 
1998, \aap, 337, 253
\reference{ark97}Arkhipova, V.P., Noskova, R.I. 1997, Astronomy Letters, 
23, 623
\reference{ark98}Arkhipova, V.P., Esipov, V.F., Noskova, R.I., Sokol, 
G.V., Tatarnikov, A.M., Shenavrin, V.I., Yudin, B.F., Munari, U., 
Rejkuba, M. 1998, Astronomy Letters, 24, 248
\reference{ark99}Arkhipova, V.P., Noskova, R.I., Esipov, V.F., Sokol, 
G.V. 1999, Astronomy Letters, 25, 615
\reference{asp97}Asplund, M. 1997, private communication
\reference{asp98}Asplund, M. 1998, \aap, 330, 641
\reference{agl97}Asplund, M., Gustafsson, B., Lambert, D.L., Rao, 
N.K. 1997, \aap, 321, L17
\reference{bar96}Barnbaum, C., Stone, R.P.S., Keenan, P.C. 1996, \apjs, 105,
419. 
\reference{bat89}Bath, G.T., Harkness, R.P. 1989, in Classical Novae,
eds. M.F. Bode and A. Evans, Chichester: J. Wiley, p. 61
\reference{bid73}Bidelman, W.P. 1973, \baas, 5, 442
\reference{blo95}Bl\"ocker, T. 1995, \aap, 299, 755
\reference{blo97}Bl\"ocker, T., Sch\"onberner, D. 1997, \aap, 324, 991
\reference{cla96}Clayton, G.C.,1996, \pasp, 108, 225
\reference{cla97}Clayton, G.C., De Marco, O. 1997, \aj, 114, 2679
\reference{cot90}Cottrell, P.L., Lawson, W.A., Buchhorn, M. 1990, 
\mnras, 244, 149 
\reference{due96a}Duerbeck, H.W., Benetti, S., 1996, \apjl, 468, 
L111 (D96)
\reference{due96b}Duerbeck, H.W., Pollacco, D., 1996, \iaucirc, 6328
\reference{due97}Duerbeck, H.W., Benetti, S., Gautschy, A., van Genderen, 
A., Kemper, C., Liller, W., Thomas, T. 1997, \aj, 114, 1657 (D97)
\reference{due98}Duerbeck, H.W., van Genderen, A., Gautschy, A., 
Pavlenko, Ya.V., Brogt, E., Janson, M., Jones, A.F., Kurk, J., Liller, 
W., Sterken, C., Voskes, T. 1998, in Unsolved Problems in Stellar 
Evolution, poster papers from the STScI Symposium, ed. M. Livio, 
May 1998, p. 17
\reference{eyr98a}Eyres, S.P.S., Evans, A., Geballe, T.R., Salama, A.,
Smalley, B. 1998, \mnras, 298, L37 
\reference{eyr98b}Eyres, S.P.S., Richards, A.M.S., Evans, A., Bode, 
M.F. 1998, \mnras, 297, 905
\reference{eyr99}Eyres, S.P.S., Smalley, B., Geballe, T.R., Evans, A.
Asplund, M., Tyne, V.H. 1999, \mnras, 307, L11
\reference{fea97}Feast, M.W., Carter, B.S., Roberts, G., Marang, F.,
Catchpole, R.M. 1997, \mnras, 285, 317
\reference{fea99}Feast, M., Whitelock, P.A. 1999, private communication
\reference{fer86}Fernie, J.D., Percy, J.R., Richer, M.G. 1986, \pasp, 
98, 605
\reference{fuh81}Fuhrmann, B. 1981, Mitt. Ver\"and. Sterne (Sonneberg), 
9, 13
\reference{gal76}Gallagher, J.S., Ney, E.P., 1976, \apj, 204, L35
\reference{gen95}van Genderen, A.M., Gautschy, A. 1995, \aap, 294, 453
\reference{gon98}Gonzalez, G., Lambert, D.L., Wallerstein, G., 
Kameswara Rao, N., Smith, V.V., McCarthy, J.K. 1998, \apjs, 114, 133
\reference{gui98}Guinan, E.F., Dewarf, L.E., McCook, G.P., Dituro, P.,
Mittal, R., Margheim, S.J., 1998, \baas, 193, 1510
\reference{hak97}Hakkila, J., Myers, J.M., Stidham, B.J., Hartmann, 
D.H. 1997, \aj, 114, 2043
\reference{har96}Harrison, T.E. 1996, \pasp, 108, 1112
\reference{hin99}Hinkle, K., Joyce, R. 1999, \iaucirc, 7266
\reference{ibe95} Iben jr., I., MacDonald, J. 1995, in White Dwarfs,
eds. D. Koester \& K. Werner, Springer, Berlin, p. 48
\reference{ive97}Ivezi\'c, Z., Nenkova, M., Elizur, M. 1997, User Manual 
for {\sc Dusty}, University of Kentucky
\reference{jac79}Jacoby, G. 1979, \pasp, 91, 754
\reference{jac99}Jacoby, G. 1999, \iaucirc, 7155
\reference{jac98a}Jacoby, G., De Marco, O. 1998, \iaucirc, 7065
\reference{jac98b}Jacoby, G., De Marco, O., Sawyer, D.G. 1998, \aj, 116, 1367
\reference{jur99}Jurcsik, J., Montesinos, B. 1999, New Astronomy Reviews, 43,
415 
\reference{kam99}Kamath, U.S., Ashok, N.M. 1999, \mnras, 302, 512
\reference{kau98}Kaeufl, H.U., Stecklum, B. 1998, \iaucirc, 6938
\reference{ker99}Kerber, F., Blommaert, J.A.D.L., Groenewegen, M.A.T.,
Kimeswenger, S., K\"aufl, H.U., Asplund, M., 1999, \aap, 350, L27 
\reference{kim97}Kimeswenger, S., Gratl, H., Kerber, F., Fouqu\'e, P., 
Kohle, S., Steele, S. 1997, \iaucirc, 6608
\reference{kim98}Kimeswenger, S., Kerber, F. 1998, \aap, 330, L41
\reference{kip99}Kipper, T. 1999, Inf. Bull. Variable Stars, 4707
\reference{kip97}Kipper, T., Klochkova, V.G. 1997, \aap, 324, L65 
\reference{law89}Lawson, W.A., Cottrell, P.L. 1989, \mnras, 240, 689
\reference{law90}Lawson, W.A., Cottrell, P.L., Kilmartin, P.M., Gilmore,
A.C. 1990, \mnras, 247, 91
\reference{law92}Lawson, W.A., Cottrell, P.L., Gilmore, A.C., Kilmartin,
P.M. 1992, \mnras, 256, 339
\reference{lil98a}Liller, W., Janson, M., Duerbeck, H.W., van Genderen, A.M.
1998a, \iaucirc, 6825
\reference{lil98b}Liller, W., Duerbeck, H.W., van der Meer, A., 
van Genderen, A.M. 1998b, \iaucirc, 7049
\reference{lyn98}Lynch, D.K., Russell, R.W., Rice, C.J., Sitko, M. 
1998, \iaucirc, 6952
\reference{mar97}Margheim, S.J., Guinan, E.F., McCook, G.P. 1997, 
\baas, 191, 4301 
\reference{mat91}Mattei, J.A., Waagen, E.O., Foster, E.G. 1991, 
R Coronae Borealis light curves: 1843 -- 1990, AAVSO Monograph No. 4, 
Cambridge, Mass.
\reference{men97}Menzies, J.W., Feast, M.W. 1997, \mnras, 285, 358
\reference{nak96}Nakano, S., Benetti, S., Duerbeck, H.W. 1996, \iaucirc, 6322
\reference{ney78}Ney, E.P., Hatfield, B.F. 1976, \apj, 219, L111
\reference{pav00}Pavlenko, Ya.V., Yakovina, L.A., Duerbeck, H.W. 
2000, \aap, in press
\reference{pol99}Pollacco, D. 1999, \mnras, 304, 127
\reference{que78}Querci, M., Querci, F. 1978, \aap, 70, L45
\reference{tak97}Takamizawa, K. 1997, VSOLJ Variable Star Bulletin, 25, 4
\reference{sei85}Seitter, W.C. 1985, Mitt. Astr. Ges., 63, 181
\reference{tyn99}Tyne, V.H., Eyres, S.P.S., Geballe, T.R., Evans, A., 
Smalley, B., Duerbeck, H.W., Asplund, M. 1999, \mnras, submitted
\reference{wer99}Werner, K., Dreizler, S., Rauch, T., Koesterke, L., Heber,
U. 1999, in IAU Symp. 191, Asymptotic Giant Branch Stars, eds. T. Le Bertre,
A. L\`ebre, and C. Waelkens (San Francisco: ASP)
\reference{woi96}Woitke, P., Goeres, A., Sedlmayer, E. 1996, \aap, 313, 217
\end{references}
\end{document}